\documentclass[12pt,a4paper,reqno]{article}
\usepackage{bm}
\usepackage{algorithm}
\usepackage{algpseudocode}
\usepackage{amsmath}
\usepackage{authblk}
\usepackage{amsfonts}
\usepackage{pgfplots} 
\usepackage[numbers,square,compress]{natbib}
\usepackage{color}
\usepackage[caption=false]{subfig}
\usepackage{float}

\usepackage{hyperref}
\hypersetup{
    unicode=false,  
    pdftoolbar=true,
    pdfmenubar=true,
    pdffitwindow=false,
    pdfstartview={FitH},
    pdftitle={Multispecies structure-preserving particle discretization of the Landau collision operator},
    pdfauthor={Eero Hirvijoki},
    pdfsubject={Gyrokinetics},
    pdfcreator={Eero Hirvijoki},
    pdfproducer={GNU Make},
    pdfkeywords={Gyrokinetics} {Maxwell's equations},
    pdfnewwindow=true,
    colorlinks=true,
    linkcolor=blue,
    citecolor=blue,
    filecolor=blue,
    urlcolor=blue
}

\newcommand{\revision}[1]{{\color{black}#1}}

\makeatother

\begin{document}

\title{Multispecies structure-preserving particle discretization of the Landau collision operator}
\author{Filippo Zonta}
\affil{Department of Applied Physics, Aalto University, P.O. Box 11100, FI-00076 AALTO, Finland}
\author{Joseph V. Pusztay}
\affil{ University at Buffalo, Buffalo,  14260, New York, USA}
\author{Eero Hirvijoki}
\affil{Department of Mechanical Engineering, Aalto University, P.O. Box 14400, FI-00076 AALTO, Finland}
\date{\today}

\maketitle
\begin{abstract}
    This paper proposes a novel numerical integrator for modeling multispecies Coulomb collisions in kinetic plasmas. The proposed scheme provides an energy-, momentum-, \revision{and positivity-}preserving particle discretization of the nonlinear Landau collision operator, extending the works of J.A. Carrillo et al., Journal of Computational Physics, 7, 100066 (2020) and E. Hirvijoki, Plasma Physics and Controlled Fusion, 63, 044003 (2021). The discrete-time conservation properties are analyzed both algebraically and numerically, and an efficient, GPU-parallelized implementation is validated against inhomogeneous temperature relaxation, isotropization and thermalization examples. The results agree with analytical estimates, \revision{confirming the method capable of reproducing physics}.
\end{abstract}

\section{Introduction}

The Vlasov--Maxwell--Landau system plays a crucial role in understanding the physics of charged particles in plasmas, especially in the important applications to fusion reactors such as the upcoming ITER experiment. It therefore should not come as a surprise that modeling the system has attracted significant interest within the computational plasma physics community. 
Achieving the target, to model the long-time behaviour of the plasma from kinetic principles accurately, 
involves, however, the challenge that the numerical scheme, after temporal and spatial discretization, ought to respect the basic properties of both the collisionless Vlasov--Maxwell equations and the dissipative Landau collision integral, as close to as possible. 

Much effort has been put in the development of structure-preserving geometric particle-in-cell schemes for the Vlasov--Maxwell part \cite{Squire-Qin-Tang-PIC:2012PhPl,Evstatiev-shadwick:2013JCoPh,Shadwick-Stamm-Evstatiev:2014PhPl,Stamm-Shadwick-Evstatiev:2014ITPS,Xiao-et-al-kinetic:2015PhPl,He-et-al-Hamiltonian-splitting:2015PhPl,Qin-et-al:2016NucFu,Xiao-et-al-fluid:2016PhPl,Kraus-et-al:2017JPlPh,Xiao-et-al:2018PlST,Xiao-Qin-6d-tokamak:2020arXiv,Hirvijoki_Kormann_Zonta:2020PoP}, which typically provide superior conservation properties by tackling the discretization of the variational or Hamiltonian structure of the system, guaranteeing that many of the geometric structures and conserved quantities are preserved at a discrete level.
Despite this success in addressing the collisionless part, a compatible structure-preserving discretization of the Landau operator has remained more elusive. 
Existing structure-preserving integrators for the Landau operator utilizing grids \cite{Yoon2014,Hager2016,Hirvijoki_Adams:2017PhPl,Adams_et_al:2017SIAM,Shiroto-Sentoku:2019PhRvE, adams2022a} use finite difference, finite volume, or finite element discretizations and, although do satisfy certain conservation laws, do not necessarily ensure positivity of the distribution function. Enforcing the positivity, on the other hand, makes it difficult to exploit the fundamental symmetry properties of the Landau operator, requiring additional measures \cite{Taitano_et_al:2015JCP,Taitano_et_al_2017JCoPh,Daniel_et_al:2020CPC}. 
The alternative to grid-based schemes, the binary collision algorithm \cite{takizukaabe:1977,Miller_Combi:1994,WANG1996,nanbu:1997,Nanbu_Yonemura:1998} that would more naturally fit the particle-in-cell method, is limited to using equal particle weights or otherwise does not guarantee energy and momentum conservation. In applications where the particle density varies significantly, equal marker-particle weights can severely limit the statistics of particle-in-cell simulations. To ensure proper statistics and resolution, schemes that admit arbitrarily weighted markers yet provide conservation laws are therefore desired.

A potential solution to this dilemma emerged only rather recently when a marker-based flow of the Landau operator, driven by the variation of an entropy functional, was shown to lead to conservation of energy and momentum, and to dissipation of entropy \cite{Carrillo-et-al:2020}. 
While perhaps exotic, the idea of interpreting diffusive phenomena as a vector field driving compressible flow is in fact straightforward. A diffusion equation $\partial_tf=\partial_i(D^{ij}\partial_jf)$
can be interpreted as an advection equation $\partial_tf+\partial_i(V^if)=0$,
where the vector field $V^i=D^{ij}\partial_j(\delta \mathcal{S}/\delta f)$
is driven by an entropy functional $\mathcal{S}=\int f\ln f$.
The analogy to incompressible flow driven by a Hamiltonian functional is clear: the Hamiltonian is replaced by the entropy and the Poisson tensor is replaced by the diffusion tensor. 
After the pioneering work \cite{Carrillo-et-al:2020}, a full discrete-time energy- and momentum-conserving and entropy-dissipating scheme was proposed in \cite{Hirvijoki2021}, revealing also the underlying metriplectic nature of the discretization to which many of the discrete structure-preserving properties can ultimately be credited to, and progress has since been made also in modeling the full Vlasov-Poisson-Landau system \cite{pusztay_dissertation}. 

In the present paper, the works \cite{Carrillo-et-al:2020,Hirvijoki2021} are further extended in order to study and account for multiple particle species. The validity of the discrete-time conservation laws and the applicability of the method to inhomogeneous multispecies collisional relaxation is demonstrated and discussed: In section \ref{multispecie-theory} the metriplectic formulation of the Landau operator is briefly reviewed and the discretization presented. Numerical tests with both electrons and ions present are then carried out in section \ref{numerical-results-si} with realistic mass ratio and compared to analytic estimates. Finally, \revision{a discussion on open questions is provided in section \ref{sec:discussion} and a summary concludes the work in section~\ref{sec:conclusion}.}

\section{Discretization of the collision operator}\label{multispecie-theory}

\subsection{From continuous phase space to marker particles}
The distribution function of charged particles in a magnetized plasma is reasonably well described by the Vlasov--Maxwell--Landau model. The Landau part, describing Coulomb collisions between the species $s$ and $\overline{s}$, is local in configuration space, affecting only the velocity part of the distribution function:
\begin{align}\label{landau-int}
    \left.\frac{\partial f_{s}}{\partial t}\right|_{\text {coll }} = -\sum_{\overline{s}} \frac{\nu_{s \overline{s}}}{m_{s}} \frac{\partial}{\partial \boldsymbol{v}} \cdot\int \delta(\boldsymbol{x}-\overline{\boldsymbol{x}}) f_{s}(\boldsymbol{z}) f_{\overline{s}}(\overline{\boldsymbol{z}}) \mathbb{Q}(\boldsymbol{v}-\overline{\boldsymbol{v}}) \cdot \boldsymbol{\Gamma}_{s \overline{s}}(\mathcal{S}, \boldsymbol{z}, \overline{\boldsymbol{z}}) d \overline{\boldsymbol{z}},
\end{align}
where $\boldsymbol{z} = (\boldsymbol{x},\boldsymbol{v})$ are the phase-space coordinates, $\nu_{s\overline{s}} = e^2_{s} e^2_{\overline{s}} \ln \Lambda_{s\overline{s}} / (8\pi\epsilon_0^2)$, the antisymmetric vector $\boldsymbol{\Gamma}$ is
\begin{align}\label{gamma}
    \boldsymbol{\Gamma}_{s \overline{s}}(\mathcal{A}, \boldsymbol{z}, \overline{\boldsymbol{z}})=\frac{1}{m_{s}} \frac{\partial}{\partial \boldsymbol{v}} \frac{\delta \mathcal{A}}{\delta f_{s}}(\boldsymbol{z})-\frac{1}{m_{\overline{s}}} \frac{\partial}{\partial \overline{\boldsymbol{v}}} \frac{\delta \mathcal{A}}{\delta f_{\overline{s}}}(\overline{\boldsymbol{z}}),
\end{align}
$\mathcal{S}$ is the entropy functional
\begin{align}\label{S}
    \mathcal{S}=-\sum_{s} \int f_{s} \ln f_{s} d \boldsymbol{z},
\end{align}
and the matrix $\mathbb{Q}(\boldsymbol{\xi})$\revision{, for any $\boldsymbol{\xi}\in\mathbb{R}^3$, is defined
\begin{align}
    \mathbb{Q}(\boldsymbol{\xi})=\frac{1}{|\boldsymbol{\xi}|}\left(\mathbb{I}-\frac{\boldsymbol{\xi} \boldsymbol{\xi}}{|\boldsymbol{\xi}|^{2}}\right).
\end{align}
}
It is straightforward to verify that the Landau operator admits a weak formulation by means of a symmetric, positive semidefinite metric bracket
\begin{align}\label{metric-bracket}
    (\mathcal{A}, \mathcal{B})=\sum_{s, \overline{s}} \frac{1}{2} \iint \boldsymbol{\Gamma}_{s \overline{s}}(\mathcal{A}, \boldsymbol{z}, \overline{\boldsymbol{z}}) \cdot \mathbb{W}_{s \overline{s}}(\boldsymbol{z}, \overline{\boldsymbol{z}}) \cdot \boldsymbol{\Gamma}_{s \overline{s}}(\mathcal{B}, \boldsymbol{z}, \overline{\boldsymbol{z}}) d \overline{\boldsymbol{z}} d \boldsymbol{z},
\end{align}
where the matrix $\mathbb{W}$ is
\begin{align} \label{W}
    \mathbb{W}_{s \overline{s}}(\boldsymbol{z}, \overline{\boldsymbol{z}})=\nu_{s \overline{s}} \delta(\boldsymbol{x}-\overline{\boldsymbol{x}}) f_{s}(\boldsymbol{z}) f_{\overline{s}}(\overline{\boldsymbol{z}}) \mathbb{Q}(\boldsymbol{v}-\overline{\boldsymbol{v}}),
\end{align}
and that the collisional evolution of a functional \revision{$\mathcal{A} = \sum_s \int A_s(\boldsymbol{z}) f_s(\boldsymbol{z}) d\boldsymbol{z}$} can be expressed in terms of the bracket \eqref{metric-bracket} as
\begin{align}\label{eq:bracket-evolution}
    \left.\frac{d \mathcal{A}}{d t}\right|_{\text {coll }}=(\mathcal{A}, \mathcal{S}).
\end{align}
\revision{In fact, using \eqref{gamma} and substituting the explicit expression for the functional $\mathcal{A}$, one may rewrite~\eqref{eq:bracket-evolution} as}
\begin{align}
    &\left.\dfrac{d}{dt}\right|_{\text {coll }} \sum_s \int A_s(\boldsymbol{z}) f_s(\boldsymbol{z}) d\boldsymbol{z}\nonumber\\
    &=\sum_{s, \bar{s}} \frac{1}{2} \iint\left(\frac{1}{m_s} \frac{\partial A_s(\boldsymbol{z})}{\partial \boldsymbol{v}}-\frac{1}{m_{\bar{s}}} \frac{\partial A_{\bar{s}}(\overline{\boldsymbol{z}})}{\partial \overline{\boldsymbol{v}}}\right) \cdot \mathbb{W}_{s \bar{s}}(\boldsymbol{z}, \overline{\boldsymbol{z}}) \cdot \boldsymbol{\Gamma}_{s \bar{s}}(\mathcal{S}, \boldsymbol{z}, \overline{\boldsymbol{z}}) d \overline{\boldsymbol{z}} d \boldsymbol{z} \nonumber \\
    &=\sum_{s, \bar{s}} \iint \frac{1}{m_s} \frac{\partial A_s(\boldsymbol{z})}{\partial \boldsymbol{v}} \cdot \mathbb{W}_{s \bar{s}}(\boldsymbol{z}, \overline{\boldsymbol{z}}) \cdot \boldsymbol{\Gamma}_{s \bar{s}}(\mathcal{S}, \boldsymbol{z}, \overline{\boldsymbol{z}}) d \overline{\boldsymbol{z}} d \boldsymbol{z} \nonumber \\
    &= -\sum_s\int A_s(\boldsymbol{z}) \sum_{\bar{s}} \frac{\nu_{s \overline{s}}}{m_{s}} \dfrac{\partial}{\partial \boldsymbol{v}} \cdot\int \mathbb{W}_{s \bar{s}}(\boldsymbol{z}, \overline{\boldsymbol{z}}) \cdot \boldsymbol{\Gamma}_{s \bar{s}}(\mathcal{S}, \boldsymbol{z}, \overline{\boldsymbol{z}})d \overline{\boldsymbol{z}} d \boldsymbol{z},
\end{align}
\revision{where the second-to-third-row move follows from the antisymmetry of $\boldsymbol{\Gamma}_{s \bar{s}}(\mathcal{S}, \boldsymbol{z}, \overline{\boldsymbol{z}})$ with respect to particle species $(s,\overline{s})$ and the labels $(\boldsymbol{z},\overline{\boldsymbol{z}})$, and from the fact that the sum is carried through all the species combinations. Choosing then $A_s(\bm{z})=\delta(\bm{z}-\bm{z}')$, the Landau operator \eqref{landau-int} is recovered. For further information regarding the matter, the reader is directed to e.g., the Refs.~\cite{Morrison:1986vw,Kraus2017,Hirvijoki2021}.}

With particle-based structure-preserving methods available for the Vlasov--Maxwell part, for compatibility reasons, it is natural to seek to apply the same approach to the collision operator. The particle density distribution function for species $s$ is therefore presented as
\begin{align}\label{f_h}
    f_{h,s}(\boldsymbol{z}) d \boldsymbol{z}= \sum^{N_s}_{p \in s} w_{p} \delta\left(\boldsymbol{x}-\boldsymbol{x}_{p}\right) \delta\left(\boldsymbol{v}-\boldsymbol{v}_{p}\right) d \boldsymbol{z},
\end{align}
where the sum over the markers $p$ is limited to the species $s$ and $w_p$ describes the weight of the marker $p$. Now, given a functional $\mathcal{A}$ of the distribution functions $f_s$, its variation with respect to the distribution $f_s$ is
\begin{align}
    \delta \mathcal{A}[\delta f_s]=\int \frac{\delta \mathcal{A}}{\delta f_s} \delta f_s d \boldsymbol{z}
\end{align}
In the case of $f_s$ being a discretization of the type \eqref{f_h}, for which the sets of \revision{variables $\boldsymbol{Z}_s = \{\boldsymbol{z}_p\}_{p\in s}$ and parametric weights $\boldsymbol{W}_s = \{w_p\}_{p \in s}$} define uniquely the distribution function, $\mathcal{A}$ can be seen as a function $A$ of the particle \revision{variables and parameters} according to
\begin{align}\label{A-coordinates}
    \mathcal{A}[\{f_{h,s}\}_s] = A(\{\boldsymbol{Z}_s, \boldsymbol{W}_s\}_s)
\end{align}
\revision{Treating the sets of weights $\boldsymbol{W}_s$ as fixed parameters and the phase-space points $\boldsymbol{Z}_s$ as the variable degrees of freedom,} the variation of $\mathcal{A}$ with respect to $f_s$ then becomes
\begin{align}
    \delta \mathcal{A}\left[\delta f_{h,s}\right]=\sum_{p \in s} w_{p}\left(\left.\nabla \frac{\delta \mathcal{A}}{\delta f_s}\right|_{\boldsymbol{z}_{p}} \cdot \delta \boldsymbol{x}_{p}+\left.\frac{\partial}{\partial \boldsymbol{v}} \frac{\delta \mathcal{A}}{\delta f_s}\right|_{z_{p}} \cdot \delta \boldsymbol{v}_{p}\right),
\end{align}
providing the following rules to discretize the functional derivatives
\begin{align}
    \left.\nabla \frac{\delta \mathcal{A}}{\delta f_s}\right|_{\boldsymbol{z}_{p}}=\frac{1}{w_{p}} \frac{\partial A(\{\boldsymbol{Z}_s, \boldsymbol{W}_s\}_s)} {\partial \boldsymbol{x}_{p}}\\
    \left.\frac{\partial}{\partial \boldsymbol{v}} \frac{\delta \mathcal{A}}{\delta f\revision{_s}}\right|_{\boldsymbol{z}_{p}}=\frac{1}{w_{p}} \frac{\partial A(\{\boldsymbol{Z}_s, \boldsymbol{W}_s\}_s)} {\partial \boldsymbol{v}_{p}}\label{functional-d-v}.
\end{align}
Substituting the discrete distributions $\{f_{h,s}\}_s$ and the above rules for transforming functional derivatives into the bracket \eqref{metric-bracket}, and replacing the strict delta function $\delta(\boldsymbol{x}_p - \boldsymbol{x}_{\overline{p}})$ in the matrix \eqref{W} with an indicator function $\mathbf{1}(p,\overline{p})$ that is one or zero depending on whether the particles $p$ and $\overline{p}$ are within \revision{the} same spatial collision cell, a finite-dimensional bracket acting on functions of the particle \revision{degrees of freedom} is obtained
\begin{align}\label{metric-bracket-discrete}
    (A, B)_h= \frac{1}{2} \sum_{s, \overline{s}} \sum_{p, \overline{p}} \boldsymbol{\Gamma}_{s \overline{s}}^h(A, p, \overline{p}) \cdot \mathbb{W}_{s \overline{s}}^h(p, \overline{p}) \cdot \boldsymbol{\Gamma}_{s \overline{s}}^h(B, p. \overline{p}).
\end{align}
\revision{The vector $\boldsymbol{\Gamma}_{s \overline{s}}^h(A, p, \overline{p})$ and the matrix $\mathbb{W}_{s \overline{s}}^h(p, \overline{p})$ in the bracket} are defined as
\begin{align}\label{gamma_h}
    \boldsymbol{\Gamma}_{s \overline{s}}^h(A, p, \overline{p}) &= \frac{1}{m_{s}w_p} \frac{\partial A}{\partial \boldsymbol{\boldsymbol{v}}_p} - \frac{1}{m_{\overline{s}}w_{\overline{p}}} \frac{\partial A}{\partial \boldsymbol{\boldsymbol{v}}_{\overline{p}}}, \\
    \mathbb{W}_{s \overline{s}}^h(p, \overline{p}) & = \nu_{s \overline{s}} \mathbf{1}(p, \overline{p}) w_p w_{\overline{p}} \mathbb{Q}(\boldsymbol{v}_p-\boldsymbol{v}_{\overline{p}}),
\end{align}
\revision{and,} in terms of the bracket, the collisional evolution of a function $A$ now becomes
\begin{align}\label{discrete-evol}
    \left.\frac{d A}{d t}\right|_{\text {coll }}=(A,S)_h.
\end{align}

\subsection{\revision{Regularization} of entropy and temporal discretization}
The numerical tests carried out in the next section will focus on the collision operator and the evolution of the distribution function in the velocity space. The spatial dependence is therefore dropped from here on, and the operator is assumed to be used within one spatial collision cell. To reduce computational demand, the tests will also be carried \revision{out} in a 2-D velocity space, instead of \revision{the} full 3-D \revision{velocity space that is left for future studies}. 

As discussed in Carrillo et al. \cite{Carrillo-et-al:2020}, the primary difficulty in applying the particle discretization lies in the details of the entropy functional: it is not computable for the distribution~\eqref{f_h}. The solution to this dilemma \revision{is} to \revision{regularize} the entropy functional, where the delta-distribution is first convoluted with a radial basis function $\psi_\epsilon$. \revision{While different choices for the shape of the function $\psi_{\epsilon}$ are possible, it was verified in \cite{Carrillo2019} that Gaussian functions centered at the particle positions and with covariance matrix equal to $\epsilon I$ represent a good balance between accuracy and speed.} The choice for $\psi_\epsilon$ in the present manuscript, reflecting the use of 2-D velocity space, is \revision{therefore}
\begin{align}\label{eq:psi_eps}
    \psi_{\epsilon_s}(\boldsymbol{v})=\frac{1}{2 \pi \epsilon_s} \exp \left(-\frac{|\boldsymbol{v}|^{2}}{2 \epsilon_s}\right),
\end{align}
where parameter $\epsilon_s$ can be different for each species $s$.
The \revision{regularized} entropy functional, evaluated with respect to the distribution \eqref{f_h}, then becomes
\begin{align}\label{S_eps}
    S_{\epsilon} = - \sum_{s} \sum_{p} \int w_p \psi_{\epsilon_s}\left(\boldsymbol{v}-\boldsymbol{v}_{p}\right) \ln \left(\sum_{p^{\prime}} w_{p^{\prime}} \psi_{\epsilon_s}\left(\boldsymbol{v}-\boldsymbol{v}_{p^{\prime}}\right)\right) d \boldsymbol{v}
\end{align}
and the derivative with respect to particle velocity needed in \eqref{gamma_h} becomes
\begin{align}\label{dSdV}
    \frac{\partial S_\epsilon}{\partial \boldsymbol{v}_p} 
    &= - \int \frac{\boldsymbol{v}-\boldsymbol{v}_p}{\epsilon_s}w_p \psi_{\epsilon_s}(\boldsymbol{v} - \boldsymbol{v}_p)  
    \left[ 1 + \ln \sum_{\overline{p}} w_{\overline{p}}\psi_{\epsilon_{\overline{s}}}(\boldsymbol{v} - \boldsymbol{v}_{\overline{p}}) \right] d\boldsymbol{v}.
\end{align}

\revision{The theoretical background behind the regularization of the entropy by means of the $\psi_\epsilon$ function is discussed in detail in the Refs.~\cite{Carrillo-et-al:2020, Carrillo2019}, specifically how it changes the non-discrete system. The primary effect of regularizing the entropy is that, although the original Landau equation is modified to some degree, the underlying metric bracket structure of the system is preserved. This can be seen to have benefits. If one would attempt regularizing the equations of motion directly, 
it might be difficult to perform that operation and guarantee conservation laws, even at the non-discrete level. Regularizing the entropy functional instead can be exploited to retain the original metric bracket structure, and the potential to obtain conservation laws that follow from it. This is discussed next.}

\revision{Both of the integrals in \eqref{S_eps} and \eqref{dSdV} can be computed numerically, e.g, with the help of a two-dimensional Gauss-Hermite quadrature that is constructed as an outer product of two one-dimensional Gauss-Hermite quadratures \cite{PresTeukVettFlan92}. For the entropy gradient, this results in:}
\begin{align}\label{discrete-entropy-gradient}
    \frac{\partial S_\epsilon}{\partial \boldsymbol{v}_p} &= - \frac{w_p \sqrt{2\epsilon_s}}{\pi\epsilon_s} \sum_{i,j} a_i a_j \boldsymbol{k}_{ij} \left[ 1 + \ln \sum_{\overline{p}} \frac{w_{\overline{p}}}{2\pi\epsilon_s} \exp\left(-\left|  \boldsymbol{k}_{ij} +  \frac{\boldsymbol{v}_p - \boldsymbol{v}_{\overline{p}}}{\sqrt{2\epsilon_s}} \right|^2 \right) \right],
\end{align}
\revision{where $\boldsymbol{k}_{ij} = (k_i, k_j)$ and $a_i$ and $a_j$ are the 1-D Gauss-Hermite weights and knots respectively.}
\revision{
}
With a discrete bracket and a computable entropy function available, the equation of motion for a marker particle is obtained by choosing $A=\boldsymbol{v}_p$ in \eqref{discrete-evol}, providing
\begin{align}\label{v-evol-2}
    \left.\frac{d \boldsymbol{v}_p}{d t}\right|_{\text {coll }} &= \frac{1}{m_s} \sum_{\overline{s}} \sum_{\overline{p}}
    w_{\overline{p}} \nu_{s \overline{s}}
    \mathbb{Q}(\boldsymbol{v}_{p}-\boldsymbol{v}_{\overline{p}})
    \cdot \boldsymbol{\Gamma}^h_{s\overline{s}}\left( S_\epsilon, p, \overline{p} \right).
\end{align}
Different ways for temporal discretization have been discussed in Ref. \cite{Hirvijoki2021}. Here, the simple momentum- and energy-conserving integrator is chosen
\begin{align}\label{time-discretization}
    \frac{\boldsymbol{v}_{p}^{n+1}-\boldsymbol{v}_{p}^{n}}{\Delta t} = \frac{1}{m_s} \sum_{\overline{s}} \sum_{\overline{p}}
    w_{\overline{p}} \nu_{s \overline{s}} 
    \mathbb{Q}(\boldsymbol{v}^{n+1 / 2}_{p}-\boldsymbol{v}^{n+1 / 2}_{\overline{p}})
    \cdot \boldsymbol{\Gamma}^{h}_{s\overline{s}}\left( S^n_\epsilon, p, \overline{p} \right),
\end{align}
where $\boldsymbol{v}^{n+1/2}_p$ is the midpoint velocity
\begin{align}
    \boldsymbol{v}_{p}^{n+1/2}=\frac{\boldsymbol{v}_{p}^{n+1}+\boldsymbol{v}_{p}^{n}}{2},
\end{align}
and $S_\epsilon^n$ refers to evaluation of the entropy with respect to time $n$. The scheme (\ref{time-discretization}) is implicit and the resulting nonlinear system of equations can be solved with, e.g., fixed-point iteration.

The discrete-time conservation laws for the simple integrator \eqref{time-discretization} are verified in a straightforward manner. 
With the discrete momentum being 
\begin{align}\label{discrete-momentum}
    P = \sum_s \sum_p w_p m_s \boldsymbol{v}_p,
\end{align}
its discrete time-rate of change is given by 
\begin{align}
    \frac{P^{n+1}-P^{n}}{\Delta t} &=
    \sum_s \sum_{p} w_{p} m_s \frac{\boldsymbol{v}_{p}^{n+1}-\boldsymbol{v}_{p}^{n}}{\Delta t}\nonumber \\
    &=\sum_{s\overline{s}}\sum_{p, \overline{p}} w_{p} \nu_{s\overline{s}}w_{\overline{p}}\mathbb{Q}\left(\boldsymbol{v}_{p}^{n+1 / 2}-\boldsymbol{v}_{\overline{p}}^{n+1 / 2}\right) \cdot \boldsymbol{\Gamma}_{s\overline{s}}\left(S_{\epsilon}^{n}, p, \overline{p}\right)\nonumber\\
    &=0.
\end{align}
The second step follows from the antisymmetry of $\boldsymbol{\Gamma}_{s\overline{s}}\left(A, p, \overline{p}\right)$ with respect to both species and particle labels. Similarly, with the discrete kinetic energy being \begin{align}\label{discrete-energy}
    K = \sum_s \sum_p \frac{w_p m_s}{2} |\boldsymbol{v}_p|^2,
\end{align}
its discrete-time evolution is given by
\begin{align}
    \frac{K^{n+1}-K^{n}}{\Delta t} &=
    \sum_s \sum_{p} w_{p} m_s \boldsymbol{v}_{p}^{n+1 / 2} \cdot \frac{\boldsymbol{v}_{p}^{n+1}-\boldsymbol{v}_{p}^{n}}{\Delta t} \nonumber\\
    &=\sum_{s\overline{s}}\sum_{p, \overline{p}} \nu_{s\overline{s}}w_{p}w_{\overline{p}}\boldsymbol{v}_{p}^{n+1 / 2} \cdot \mathbb{Q}\left(\boldsymbol{v}_{p}^{n+1/2}-\boldsymbol{v}_{\overline{p}}^{n+1/2}\right) \cdot \boldsymbol{\Gamma}_{s\overline{s}}\left(S_{\epsilon}^{n}, p, \overline{p}\right)\nonumber\\
    &=\frac{1}{2}\sum_{s\overline{s}}\sum_{p, \overline{p}} \nu_{s\overline{s}}w_{p}w_{\overline{p}}(\boldsymbol{v}_{p}^{n+1/2}-\boldsymbol{v}_{p}^{n+1/2}) \cdot \mathbb{Q}\left(\boldsymbol{v}_{p}^{n+1/2}-\boldsymbol{v}_{\overline{p}}^{n+1/2}\right) \cdot \boldsymbol{\Gamma}_{s\overline{s}}\left(S_{\epsilon}^{n}, p, \overline{p}\right)\nonumber\\
    &=0.
\end{align}
Here the second-to-last step follows again from the antisymmetry of $\boldsymbol{\Gamma}_{s\overline{s}}\left(A, p, \overline{p}\right)$ and the last from the identiy $\boldsymbol{\xi}\cdot\mathbb{Q}(\boldsymbol{\xi})=\boldsymbol{0}$.

\section{Numerical tests}\label{numerical-results-si}

In what follows, three numerical tests for the scheme \eqref{time-discretization} are presented. The aim is to study how well the integrator manages the conservation laws that are expected from the theoretical analysis, i.e., the exact conservation of energy and momentum. Furthermore, the scheme is tested against temperature and velocity isotropization to compare with analytical estimates.
The discussion is limited to a two-dimensional velocity space, leaving a more general three-dimensional integrator for future studies. 

\revision{In all of the tests, the temperatures of the species $s$ are computed from
\begin{align}
    T_{s}^x&=\dfrac{m_s}{n_s k_B}  \sum_{p \in s} w_p (v_p^x - u_s^x)^2,\\
    T_{s}^y&=\dfrac{m_s}{n_s k_B}  \sum_{p \in s} w_p (v_p^y - u_s^y)^2,\\
    \label{discrete-temperature}
    T_s &= \dfrac{m_s}{2 n_s k_B}  \sum_{p \in s} w_p | \boldsymbol{v}_p - \boldsymbol{u}_s |^2=\frac{T_s^x+T_s^y}{2},
\end{align}
where} the species density $n_s = \sum_s w_p$ is the sum of the markers weights and $\boldsymbol{u}_s$ is the species $s$ flow velocity
\begin{align}
    \boldsymbol{u}_s = \dfrac{1}{n_s} \sum_{p \in s} w_p \boldsymbol{v}_s.
\end{align}
The distribution functions are initially set to anisotropic Maxwellians
\begin{align}\label{f_anisotropic}
    f_s(\boldsymbol{v}, t=0)&=\frac{n_s m_s}{2 \pi k_B \sqrt{T_s^x T_s^y}}\left[\exp \left(-\frac{m_s\left(v^x-u_{s}^x\right)^{2}}{2 k_b T_s^x} -\frac{m_s\left(v^y-u_{s}^y\right)^{2}}{2 k_b T_s^y}\right)\right],
\end{align}
with different values for the characteristic parameters in different tests (given later). 
The weights $w_p$ are computed from $f_s$ and normalized to $n_s$, regardless of the number of markers $N_s$, according to
\begin{align}\label{w_derivation}
     w_{p \in s} = \left(\dfrac{2L}{N_s}\right)^2 f_s(\boldsymbol{v}_p, t=0).
\end{align}
\revision{As the species thermal velocities $v_{t,s} = \sqrt{k_B T_s / m_s}$ may differ substantially from each other, due to different masses, the} markers for each species are initially placed in different $N \times N$ rectangular meshes in the domain $[-L_s, L_s] \times [-L_s,L_s]$ with $L_s = 5 v_{t,s}$.

\revision{Different convolution parameters $\epsilon_s$ are chosen for each species. On one hand, it is preferable to choose small values in order to have distribution functions approaching their exact form \eqref{f_h}. On the other hand, it was proven in \cite{Carrillo2019} that these kind of regularized particle methods converge typically with $2L/N = \mathcal{O}(\epsilon)$. Small values for $\epsilon_s$ therefore call for a small average separation between markers, increasing the total number of markers and the global computational effort. At this time, theoretical estimates regarding the optimal values for $\epsilon_s$ are not available and remain an open question. For this reason, the parameter are, for now, chosen with trial and error.} The choices $\epsilon_s = 1.2 (L_s/N)^{1.98}$ have been experimentally verified to be a good compromise between \revision{speed} and convergence, similarly to the choice made in \cite{Carrillo-et-al:2020}. 

\revision{The time step for the simulations is chosen with the electron-electron collision time in mind, representing the fastest dynamics within the system and being of the order of $5\times 10^{-6}s$ for the test cases considered. The choice of $\Delta t = 1\times 10^{-8}s$ is therefore reasonable. It has also been checked numerically that a longer time step could be possible, but the value chosen happens to be appropriate for all numerical tests considered here. The markers are subsequently evolved in time with Eq.~\eqref{time-discretization} using a fixed-point iteration method, which usually requires $5$ to $10$ iterations to converge to machine precision.
}

\subsection{Electron-positron collisions} 
\begin{figure}[!htb]
    \subfloat[]{
        \includegraphics[width=0.49\textwidth]{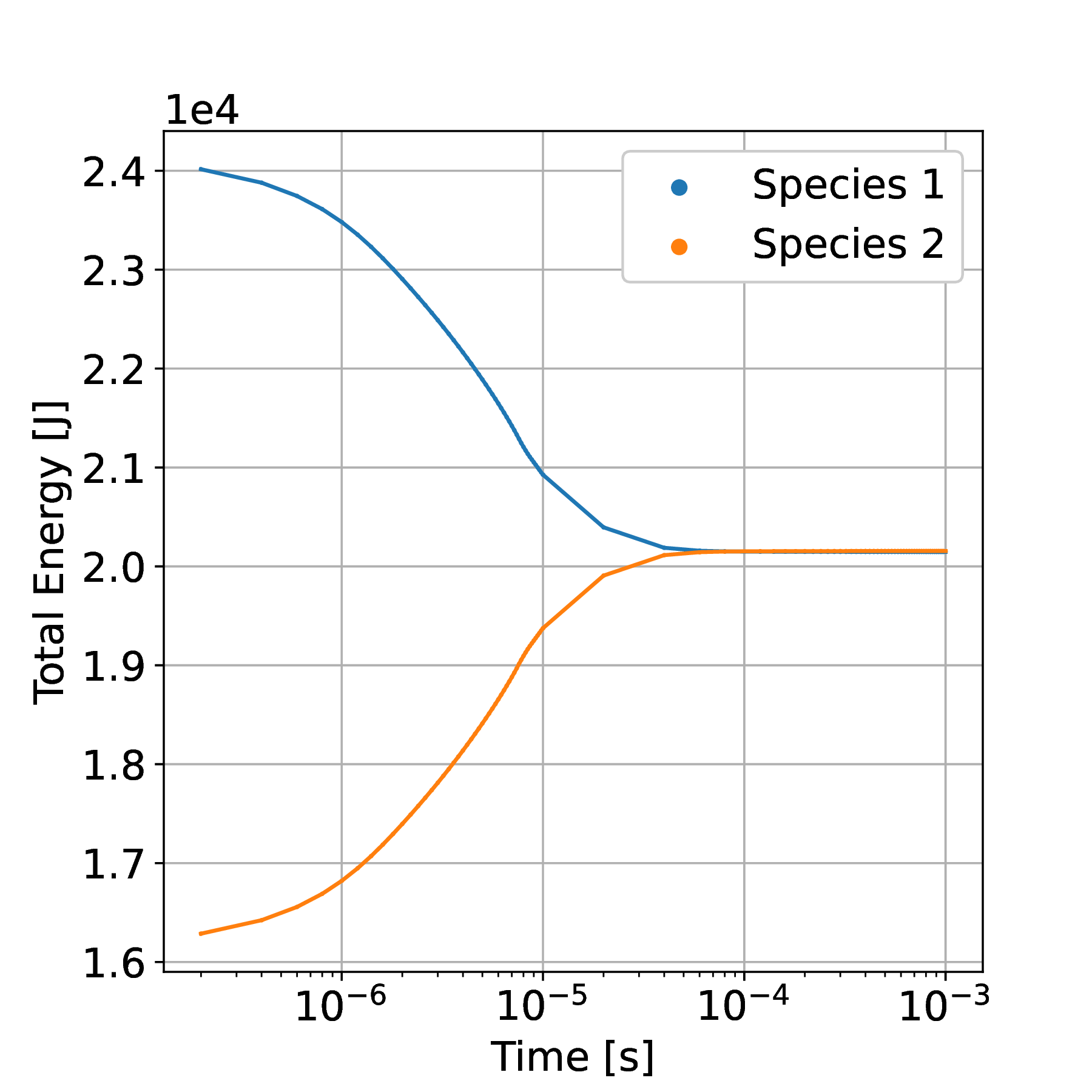}
    }
    \subfloat[]{
        \includegraphics[width=0.49\textwidth]{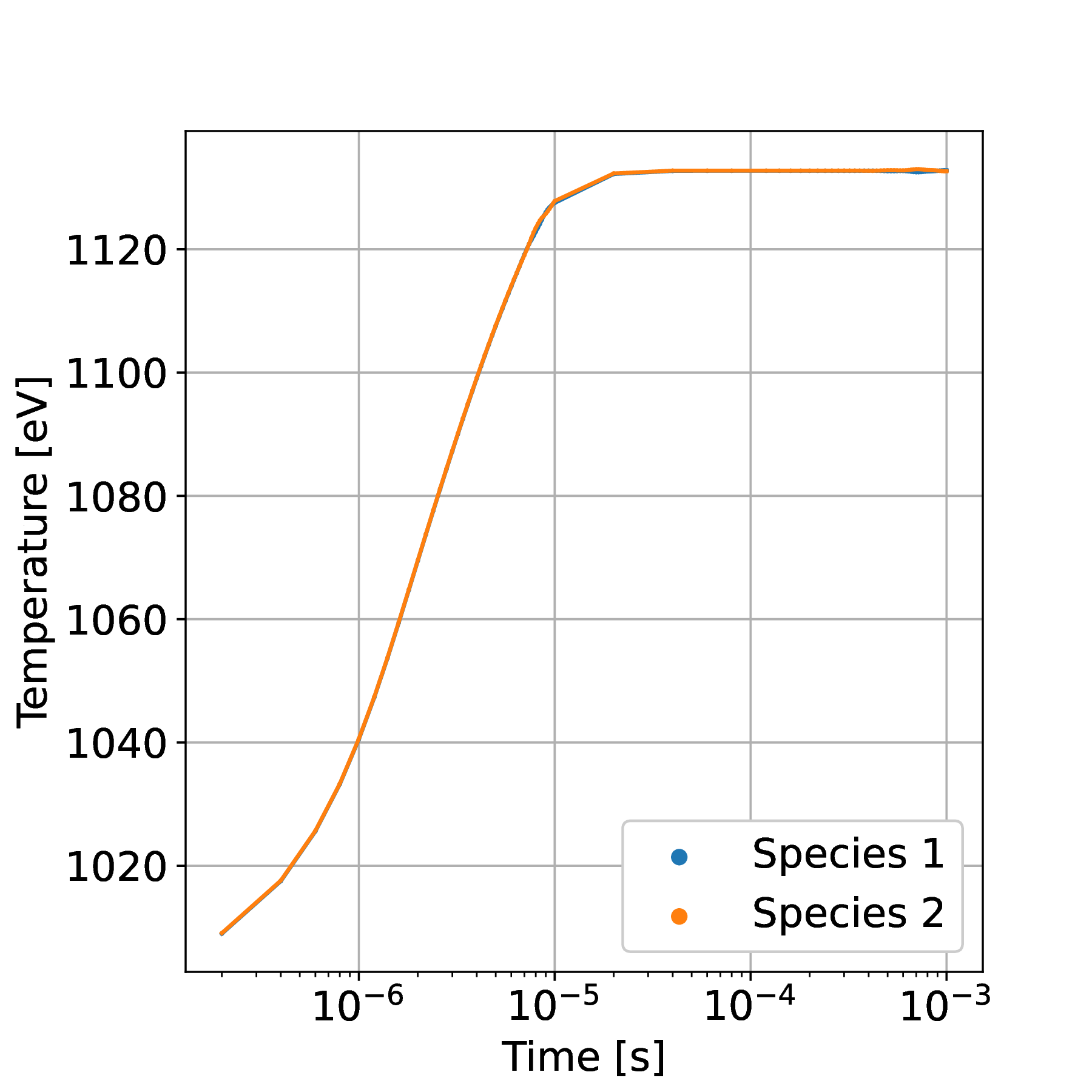}
    }\\
    \subfloat[]{
        \includegraphics[width=0.49\textwidth]{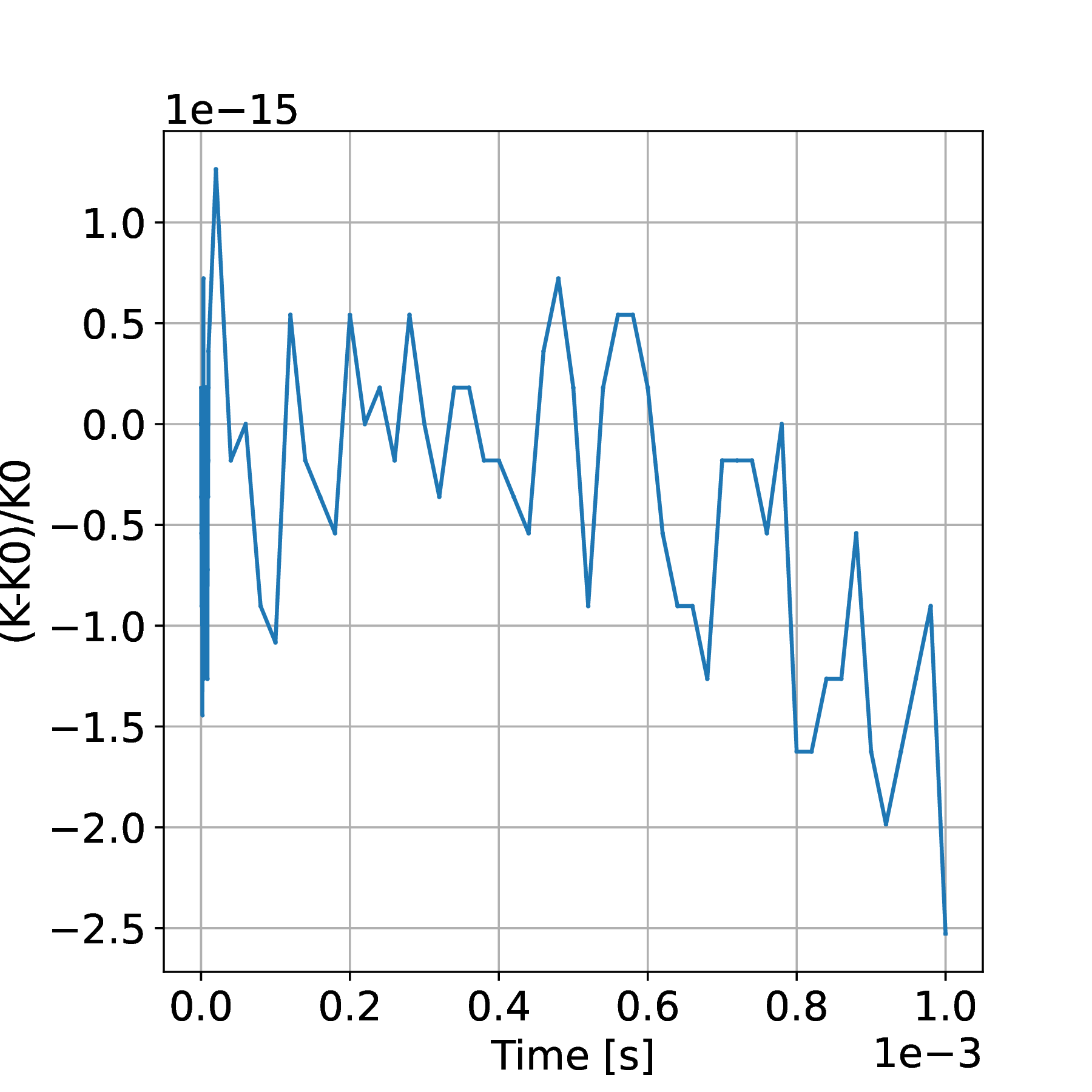}
    }
    \subfloat[]{
        \includegraphics[width=0.49\textwidth]{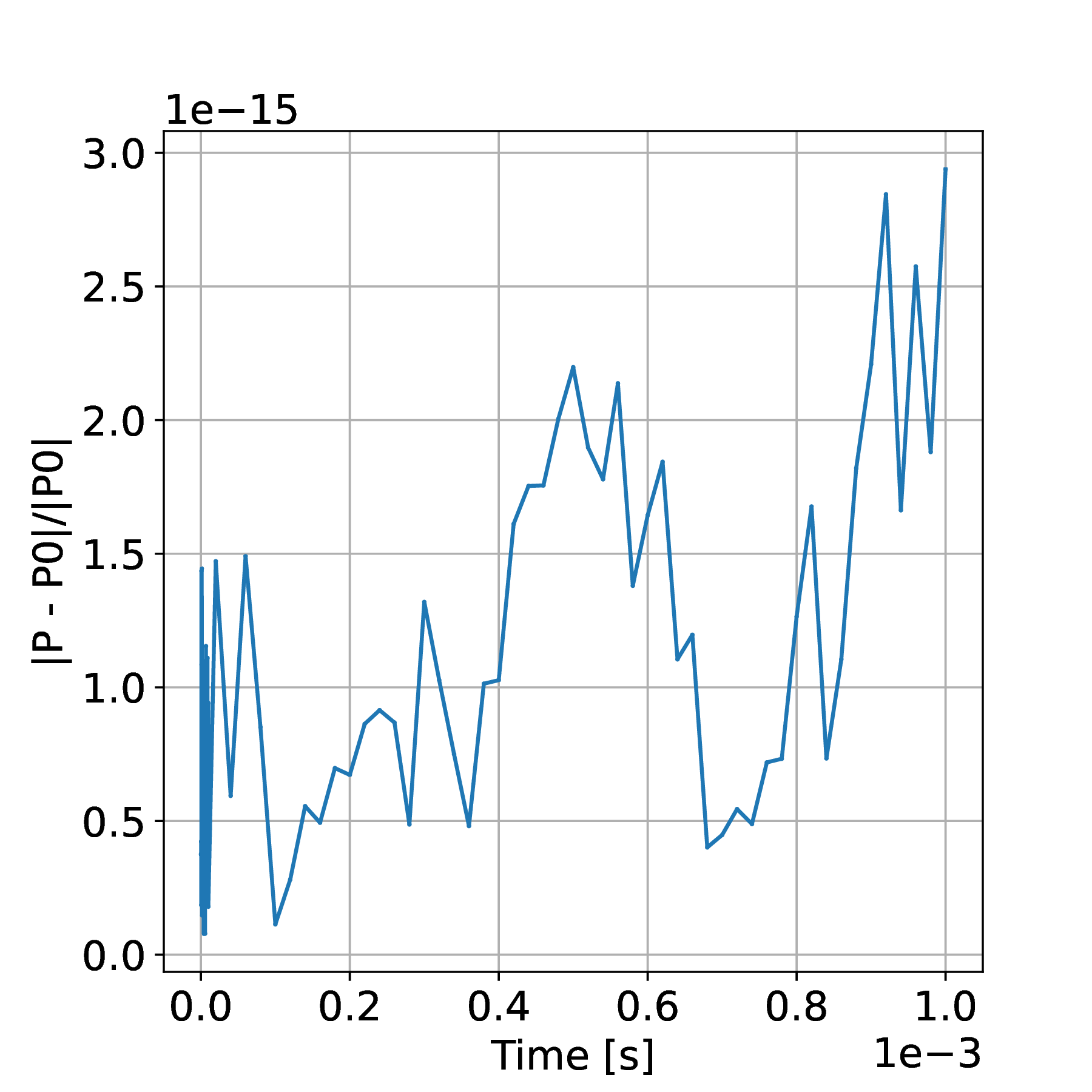}
    }
    \caption{Species energy (a) and temperature (b), total energy error (c) and momentum error (d) for an energy relaxation simulation of two species of same mass $m_e$ and an isotropic Maxwellian with $T=1\,\text{keV}$ as initial distribution. \revision{The temperature shift agrees well with the analytical estimate $\Delta T(\infty) = 125\,\text{eV}$}}
    \label{ee-relaxation}
\end{figure}

The first example is aimed at testing basic stability properties of the integrator. The masses of the two species have been set to the electron mass $m_e$, the initial distributions are isotropic Maxwellians with an equal temperature and density of $1\,\text{KeV}$ and $n=1\times10^{20}\,\text{m}^{-3}$, respectively, and the flow velocities are $u_{1,x} = -v_t$, $u_{1,y} = v_t/8$, $u_{2,x} = 0$ and $u_{2,y} = -v_t/8$, with $v_t = \sqrt{k_B T / m_e}$ the thermal speed. With these parameters the plasma is characterized by a Coulomb logarithm of approximately $\ln \Lambda = 16$. Figure~\ref{ee-relaxation} illustrates the results for a simulation of \revision{total time $t_\text{max}=1\times10^{-3}\,\text{s}$}. The two populations remain effectively thermalized for the entire simulation time while their energies converge to a half way value, as expected.

\revision{The temperature evolution is what is expected from the theory. In fact, assuming equal masses for the species, the total energy and momentum conservation implies that the energy and flows of the species, for example the first one, converge to the midpoint value:
\begin{align}
    K_1(\infty) &= \frac{1}{2} (K_1(0) + K_2(0)), \\
    \boldsymbol{u}_1(t) &= \frac{1}{2} (\boldsymbol{u}_1(0) + \boldsymbol{u}_2(0)).
\end{align}
Assuming the species to be thermalized at $t=0$, the temperature is expected to be shifted by
\begin{align}
    \Delta T_1(\infty) = \frac{m}{8k_b} \left| \boldsymbol{u}_1(0) - \boldsymbol{u}_2(0) \right|^2,
\end{align}
which translates to $125\,\text{eV}$ given the initial data. This is in good agreement with the numerical results shown in the panel (b) in Fig.~\ref{ee-relaxation}, where we have taken into account the initial numerical temperature of the particle population being slightly different from that of the smooth analytical initial distribution due to sampling.}

\subsection{Electron-ion velocity relaxation}\label{subsec:electron-ion-velocity-relaxation}
\begin{figure}[!htb]
    \subfloat[]{
        \includegraphics[width=0.49\textwidth]{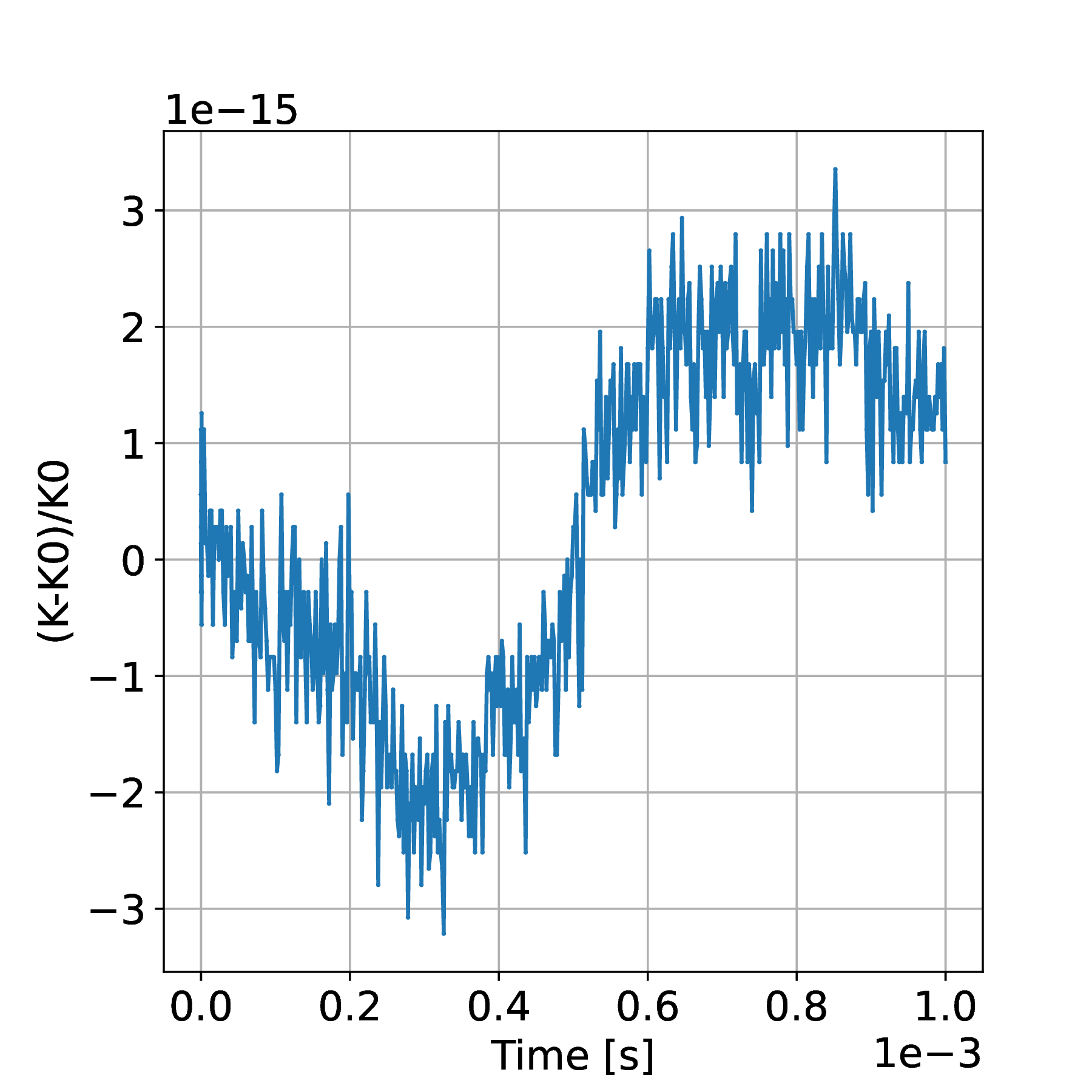}
    }
    \subfloat[]{
        \includegraphics[width=0.49\textwidth]{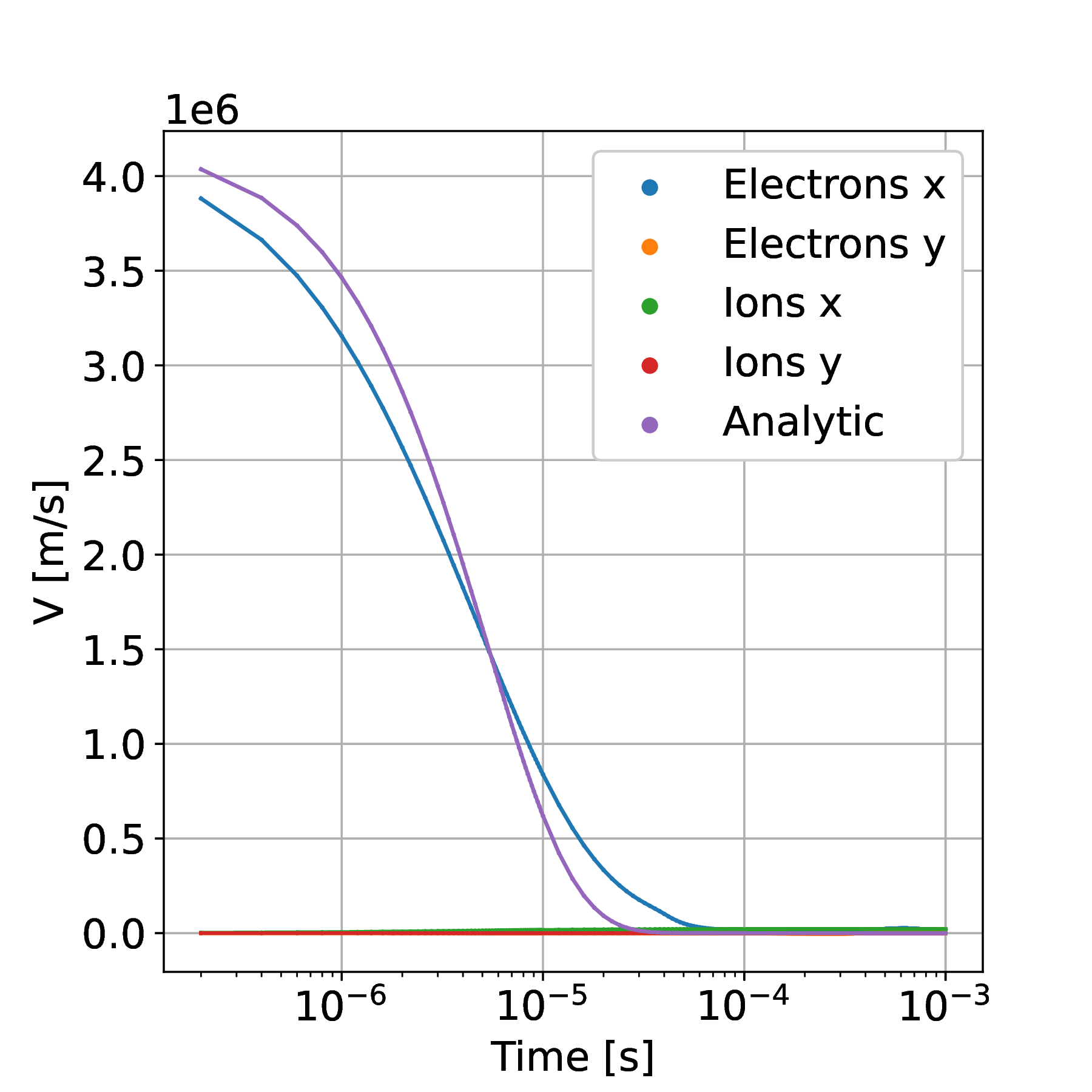}
    }\\
    \subfloat[]{
        \includegraphics[width=0.49\textwidth]{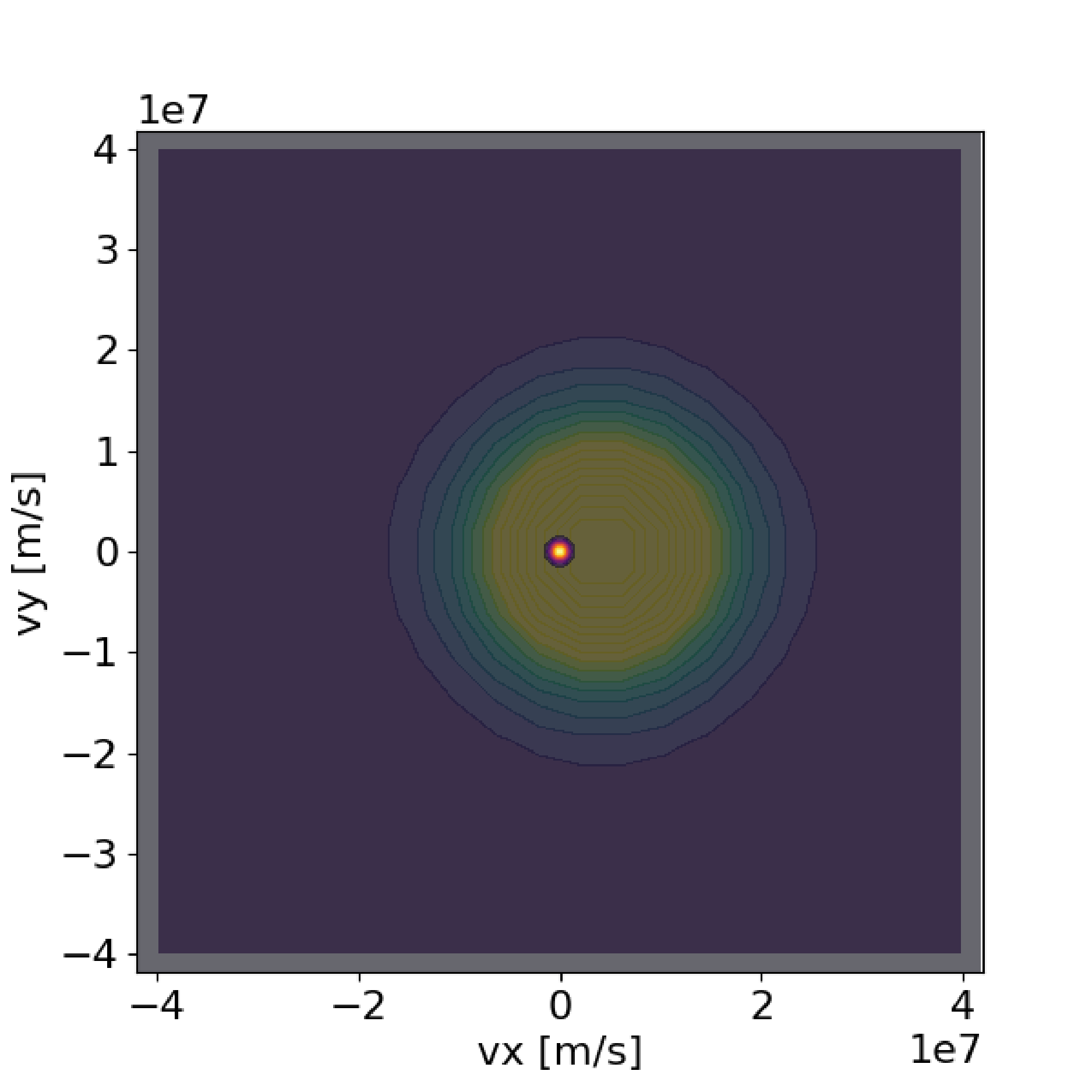}
    }
    \subfloat[]{
        \includegraphics[width=0.49\textwidth]{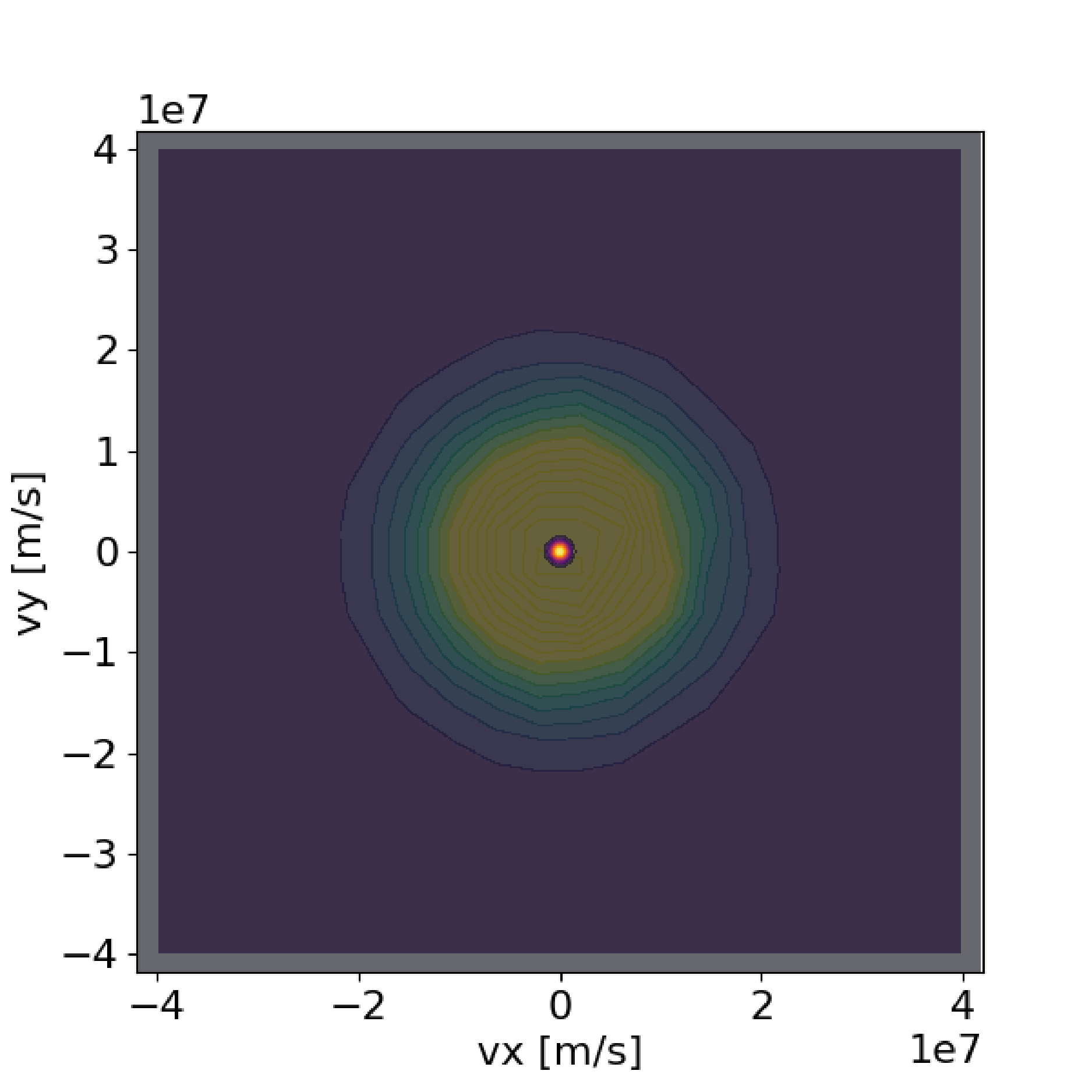}
    }
    \caption{Energy error (a) and species flow velocity (b) for a velocity relaxation simulation with masses $m_e$ and $m_i=200\,m_e$ with initially isotropic Maxwellian with temperature $T=400\,\text{eV}$ and $u_{e,x} = 2.1\times10^{6}\,\text{m/s}$. The electrons velocity \revision{relaxation agrees well with the analytical estimates $\tau_r^{e/i} = 5.2\times10^{-6}\,\text{s}$.} \revision{The distribution functions of the two species are presented at time $t=0$ (c) and $t=1\times10^{-3}\,\text{s}$ (d).}}
    \label{ei-flow-relaxation}
\end{figure}

The second example, illustrated in Fig.~\ref{ei-flow-relaxation}, is aimed at studying the relaxation of the species flow velocities. The ion mass has been set to $m_i=200\,m_e$, and the initial flow velocities are $\boldsymbol{u}_e = (v_{t,e}/2, 0)$ and $\boldsymbol{u}_i = (0, 0)$. The initial temperatures are isotropic, with $T=400\,\text{eV}$ for both species, and the density is set to $n=1\times10^{20}\,\text{m}^{-3}$. The Coulomb logarithm characterising the plasma is approximately $\ln\Lambda=15$.
The total energy and \revision{momentum} are exactly conserved at all times, as in the previous example. The electron flow velocity relaxes exponentially towards the ion flow, as we expect from a typical electron-ion relaxation process. Moreover, the relaxation time is in good agreement with the conventional estimate of the electron-ion momentum exchange theory~\cite{Spitzer1963, Hinton1976}. \revision{The analytical time-scale estimate of
\begin{align}
    \dfrac{1}{\tau_r^{e/i}} &= \frac{16}{3} \frac{\sqrt{\pi} e^4 \ln \Lambda}{\left(4 \pi \varepsilon_0\right)^2 \sqrt{m_e} \left(k_b T_e\right)^{3 / 2}},
\end{align}
for the exponential relaxation, fits well the numerical simulation.}

\subsection{Multispecies temperature isotropization with realistic mass ratio} 
In the last example, a simulation of electrons and deuterons with a real mass ratio is presented. In this case, the two species have been initialized to anisotropic Maxwellians with temperatures $T_{e}^x = 400\,\text{eV}$, $T_{e}^y = 300\,\text{eV}$, $T_{i}^x = 250\,\text{eV}$ and $T_{i}^y = 200\,\text{eV}$. The density is set to $n=1\times10^{20}\,\text{m}^{-3}$ and the Coulomb logarithm characterising the plasma is approximately $\ln \Lambda = 14.6$. 

\revision{In a 3-D case, the relaxation process is described and documented in literature. The differential equations describing the evolution of temperature along ($||$) and perpendicular to ($\perp$) a chosen coordinate direction in the velocity space are \cite{NRLFormulary, Spitzer1963, Ichimaru1970, Hinton1976}
\begin{align}
    \frac{d T_{s\perp}}{d t}=-\frac{1}{2} \frac{d T_{s||}}{d t}=-\frac{T_{s\perp}-T_{s||}}{\tau_s},
\end{align}
where the relaxation time $\tau_{s}$ is
\begin{align}\label{eq:tau_alpha}
    \tau_{s}^{-1}=\frac{2 \sqrt{\pi} e_s^2e_{\overline{s}}^2 n_s \ln\Lambda_{s\overline{s}}}{\left(4\pi\epsilon_0\right)^2 \sqrt{m_s}\left(k_b T_{||}\right)^{3 / 2}} A^{-2}\left[-3+(A+3) \frac{\tan ^{-1}\left(A^{1 / 2}\right)}{A^{1 / 2}}\right], 
\end{align}
for species $s$ and the coefficient is $A=T_\perp/T_\parallel-1$ for $T_\perp>T_\parallel$.
The rates for the parallel and perpendicular direction differ due to the fact that the perpendicular direction has two degrees of freedom while the parallel only has one. In 2-D setting, the temperatures related to $x$ and $y$ directions are both associated with only one degree of freedom, and the relaxation equations become
\begin{align}
    \frac{d T_{s}^x}{d t}&=-\frac{T_{s}^x-T_{s}^y}{\tau_s},\\
    \frac{d T_{s}^y}{d t}&=-\frac{T_{s}^y-T_{s}^x}{\tau_s},
\end{align}
with $\tau_s$ being given by \eqref{eq:tau_alpha}, with $A \equiv T_{>}/T_{<}-1$ and the $T_\parallel$ in the denominator replaced by $T_{<}$,  with $T_{>}$, $T_{<}$ the higher and lower of the temperatures $T_{s}^x$ and $T_{s}^y$ respectively. The thermal equilibration between species $s$ and $\overline{s}$ is modeled via
\begin{align}
    \frac{dT_s}{dt}=-\frac{T_s-T_{\overline{s}}}{\overline{\tau}_\epsilon^{s\backslash\overline{s}}},
\end{align}
with the thermalization time being~\cite{Hinton1976, Spitzer1963}
\begin{align}
    \dfrac{1}{\bar{\tau}_\epsilon^{s\backslash\overline{s}}} &= \dfrac{8}{3} \dfrac{ \sqrt{2\pi m_s m_{\overline{s}}} e_s^2 e_{\overline{s}}^2 n_s \ln\Lambda_{s\overline{s}}}{(4\pi\epsilon_0)^2 \left(m_s k_b T_{\overline{s}}+m_{\overline{s}}k_b T_s\right)^{3 / 2}},
\end{align}
and the species mean temperature in the 2D setting being $T_s=(T_{s}^x+T_{s}^y)/2$.}

\begin{figure}[!htb]
    \subfloat[]{
        \includegraphics[width=0.49\textwidth]{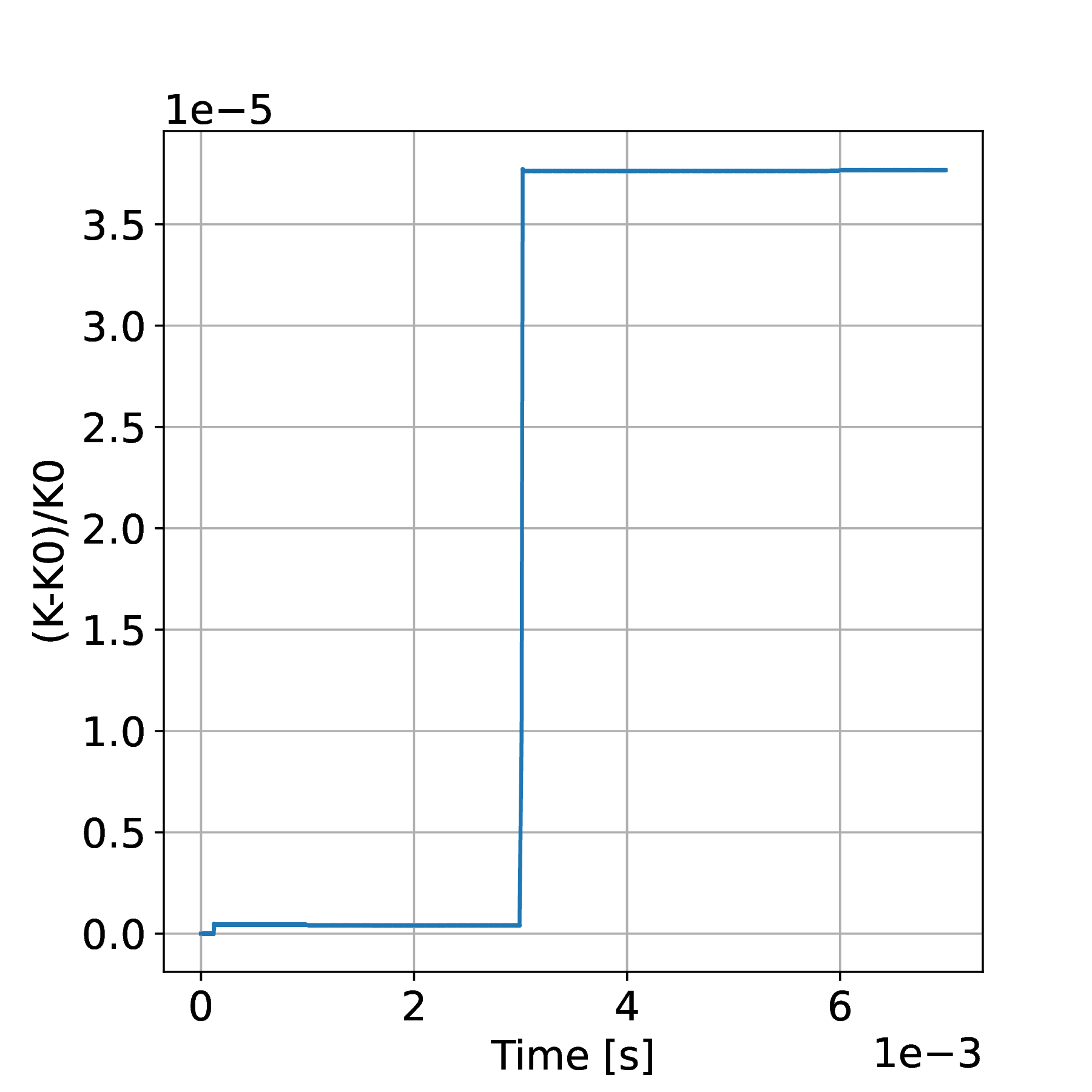}
    }
    \subfloat[]{
        \includegraphics[width=0.49\textwidth]{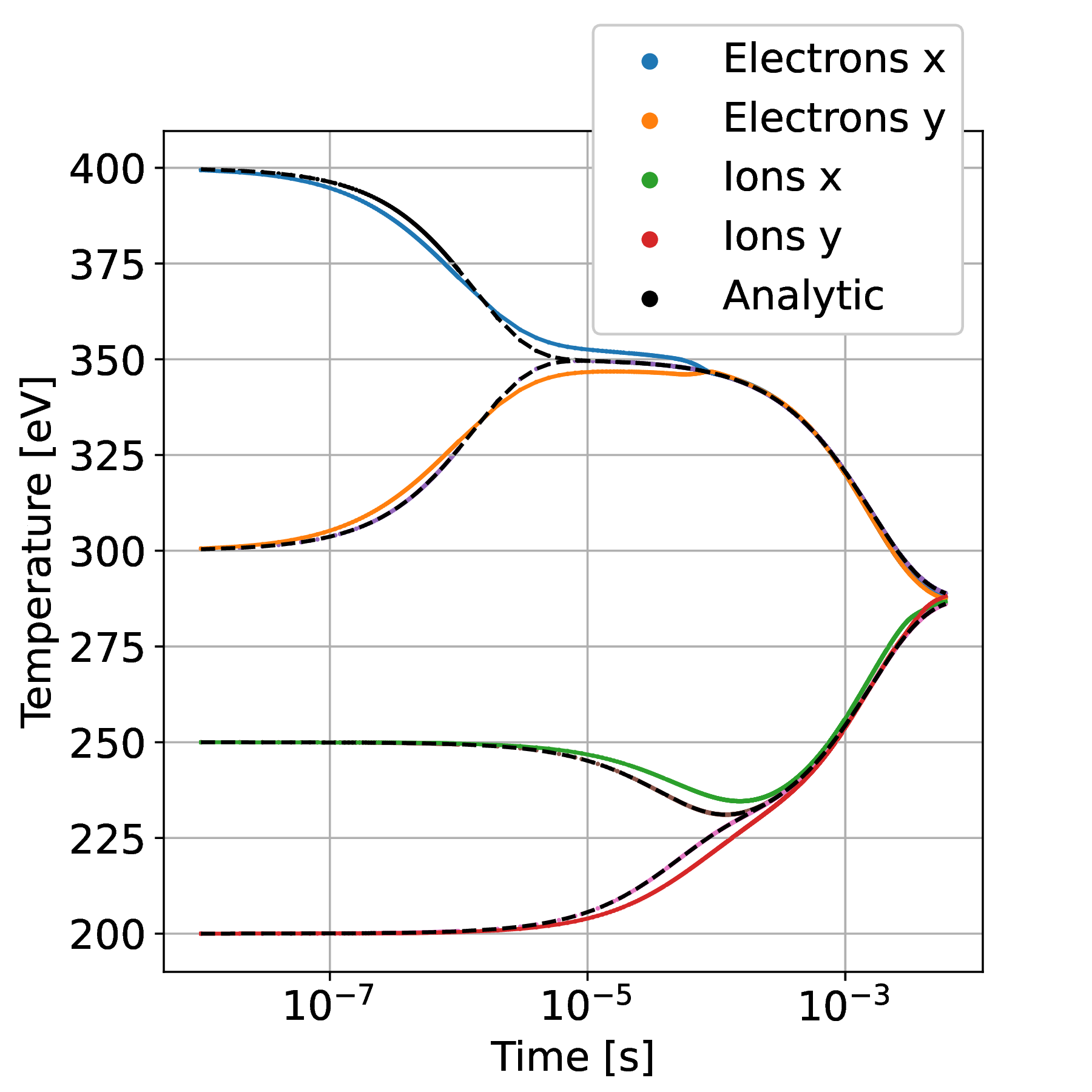}
    }\\
    \subfloat[]{
        \includegraphics[width=0.49\textwidth]{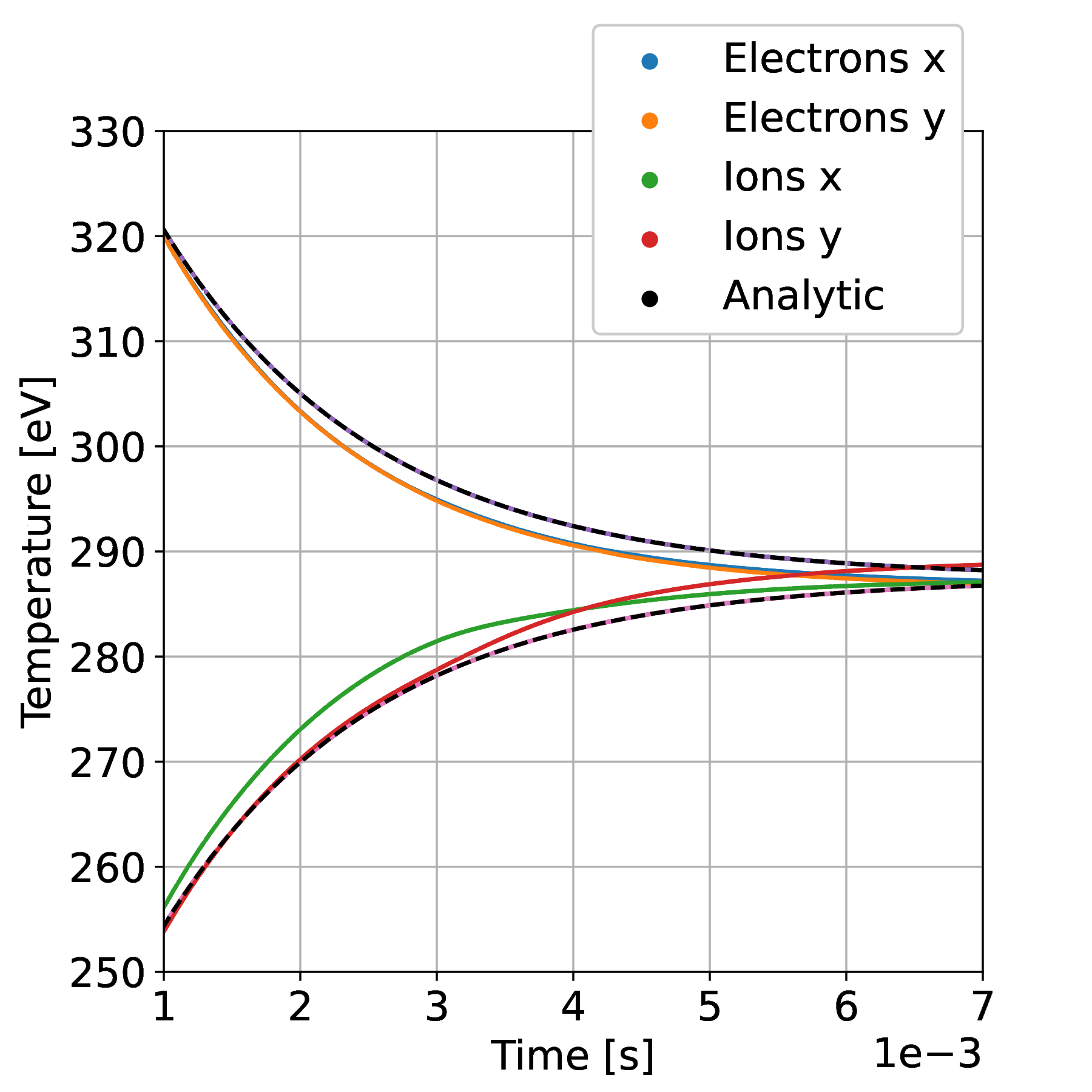}
    }
    \caption{Energy error (a) and species temperatures (b) for an isotropization and thermalization simulation of electrons and deuterons with an anisotropic Maxwellians as initial distribution. \revision{A zoomed-in view is provided in (c)}. Both the isotropization and thermalization are in good agreement with the analytical estimates \revision{$\tau_e = 5.2\times10^{-6}\,\text{s}$, $\tau_i = 1.7\times10^{-4}\,\text{s}$ and $\bar{\tau}_\epsilon^{e/i} = 1.6\times10^{-3}\,\text{s}$.}}
    \label{ei-thermal-isotrop}
\end{figure}

As evident from Fig.~\ref{ei-thermal-isotrop}, both isotropization times, $\tau_{i}$ and $\tau_e$, and the thermalization time $\overline{\tau}_\epsilon^{e\backslash i}$ are in good agreement with the analytical theory: the simulated curves match well with the analytical ones. In this example, though, while the total momentum is conserved to machine precision throughout the entire simulation, the total energy error shows a few distinctive jumps occurring. This results from the implicit system of equations for advancing the particle velocities \eqref{time-discretization} not been solved to machine precision as the solver is unable to fully converge on occasion. A root cause for this behaviour has not been identified yet but it could be related to the mass ratio as the previous examples displayed no such issue. 

\section{Discussion}\label{sec:discussion}

The method proposed in this paper involves computing effectively an $\mathcal{O}(N^2)$ algorithm in each spatial collision cell and an efficient parallelization per each collision cell is elementary. The code, with which the example calculations were performed, is therefore implemented in C++ using the open source Eigen library and CUDA for GPU parallelization. 
\revision{The execution of the code follows the steps described in Algorithm~\ref{alg:execution_description}.}
\begin{algorithm}[H]
\caption{Execution flow for the multi-species method.}
\label{alg:execution_description}
\begin{algorithmic}
\For {$s$\text{ in species}}
\State \text{Initialize markers in a rectangular mesh with parameters $L_s$, $\epsilon_s$ and $w_s$ as in \eqref{w_derivation}}
\EndFor
\While{$t < t_\text{max}$}
\State \text{Compute $\partial S / \partial \boldsymbol{v}_p$ using \eqref{discrete-entropy-gradient} assigning each component to a separate CUDA thread.}
\State Compute new $\boldsymbol{v}_p$ in a CUDA kernel solving \eqref{time-discretization} with fixed-point iterations until convergence.
\State $t \gets t + \Delta t$
\EndWhile
\end{algorithmic}
\end{algorithm}

\begin{figure}[!htb]
    \centering
    \includegraphics[width=0.5\textwidth]{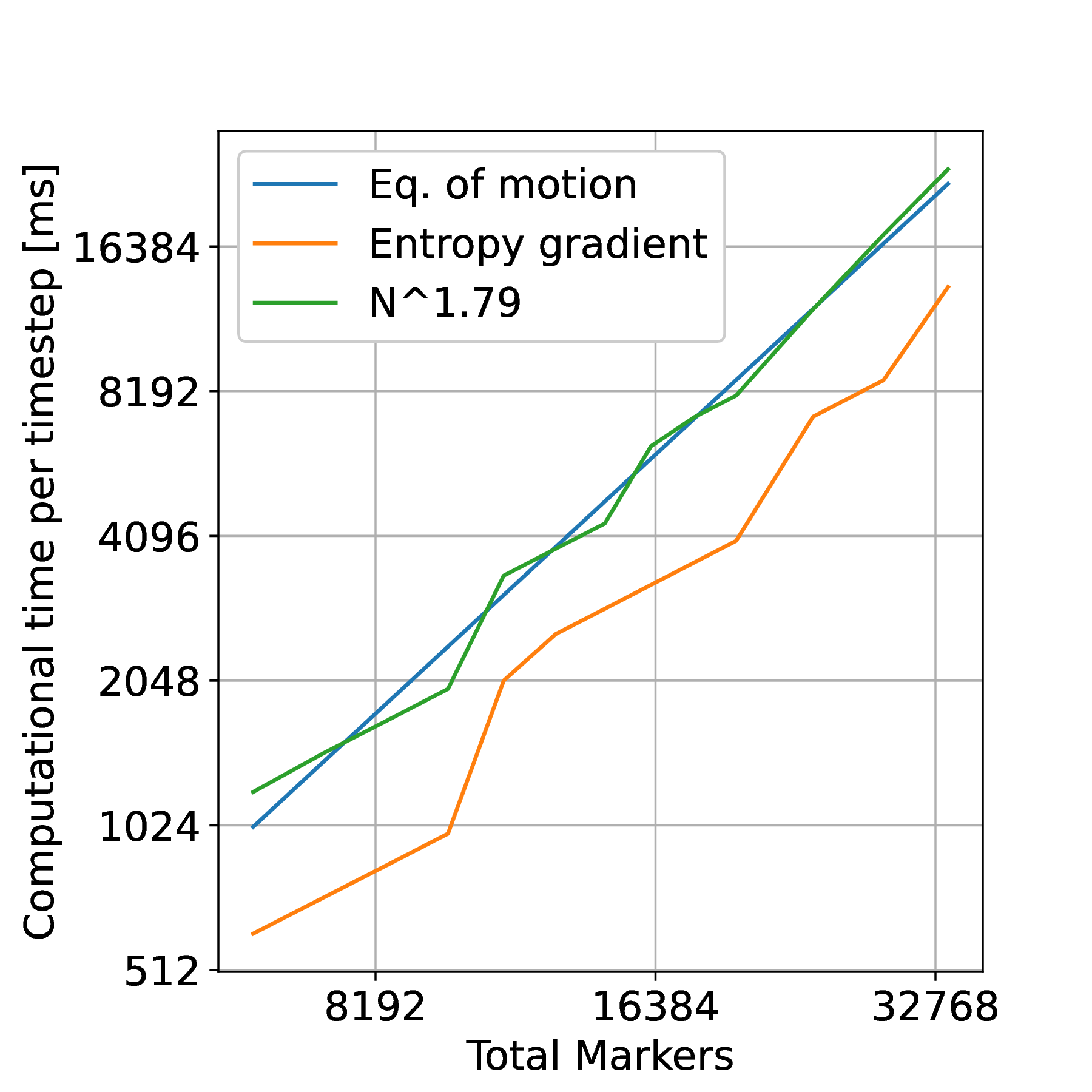}
    \caption{Time step computing time of the entropy gradient step and the fixed point iterations, plotted against the total number of markers for each species and for each dimension. The benchmark run is performed on a NVIDIA Quadro K2200 GPU.}
    \label{fig:benchmark}
\end{figure}

There are two main computationally intensive tasks. The evaluation of the discrete entropy gradient \eqref{discrete-entropy-gradient}, for which it is necessary to compute all interaction pairs between the markers, and the fixed-point iterations to solve the discrete equations of motion \eqref{time-discretization}, each with $N$ matrix-vector multiplications between the precomputed entropy gradient and the projection operator $\mathbb{Q}$. \revision{The energy, momentum and temperature of the system are not evaluated at \revision{iteration} step and therefore do not contribute substantially to the global computational time.}
Fig.~\ref{fig:benchmark} shows how the computing time for both of the two \revision{intensive tasks scales with the total number of markers. The overall scaling for both is similar and is approximately $\mathcal{O}(N^{1.8})$ for the practical cases studied here. For comparison, the binary collision scheme \cite{takizukaabe:1977} scales approximately as $\mathcal{O}(N)$ but at the cost of losing the exact conservation properties if the marker weights are non-uniform.}
As GPUs continue to develop, we anticipate that addressing also the full 3-D velocity space with the presented algorithm is feasible in the near future and could be coupled to structure-preserving particle-in-cell algorithms in a compatible manner.

The tests presented demonstrate the potential of the algorithm to provide physically meaningful results. However, there are a few issues that may occur in specific cases and are currently under investigation. First, the energy is guaranteed to be conserved to machine precision whenever the discrete equations of motion \eqref{time-discretization} are solved to machine precision. 
Although this usually happens in five or six fixed-point iterations, for some test cases and marker states the solution of the equations of motion fails to converge \revision{on occasion}, even with more refined convergence methods and regardless of the number of iterations used. This was observed in the realistic mass-ratio test case in Fig.~\ref{ei-thermal-isotrop}. 
While the root cause of the problem is currently under investigation, for now, it is useful to note that this issue seems to appear only a few times during a long simulation and that the energy jumps change the total energy by a small fraction that is within reasonable limits over the time periods investigated. The total momentum is conserved to machine precision independently, whether or not the solution converges.

\begin{figure}[!htb]
    \centering
        \includegraphics[width=0.45\textwidth]{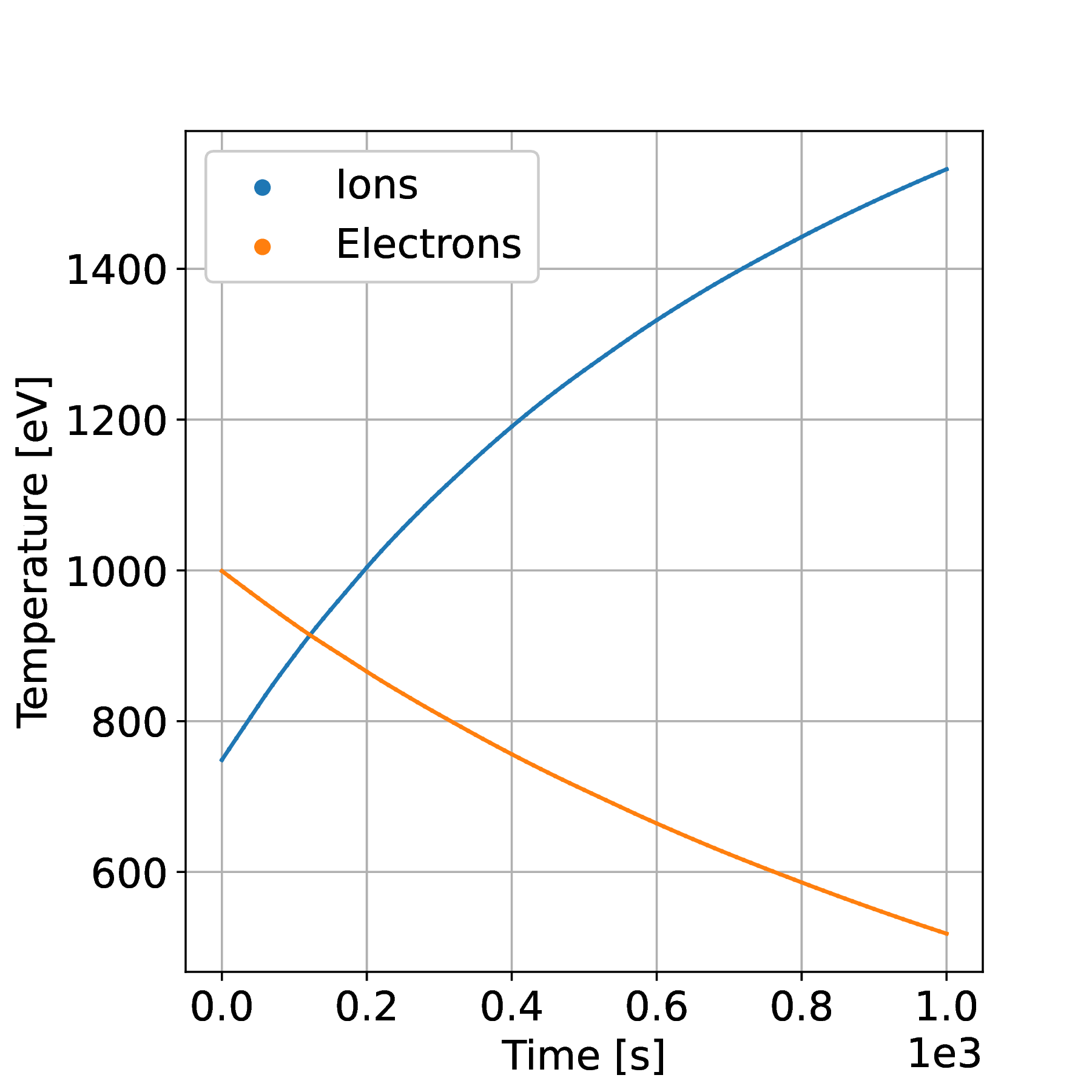}
    \caption{
    An example of a failure in a temperature relaxation case for a specific setting.}
    \label{fig:energy-jump-temperature-wrong}
\end{figure}

As the second matter, it appears that the radial-basis-function parameter $\epsilon_s$ has to be chosen carefully. In principle, $\epsilon$ should be chosen close to zero. However, \cite{Carrillo-et-al:2020} found heuristically that for a single species case, $\epsilon=(2L/N)^{0.98}$ is a practical lower limit. For the case of multiple species, $\epsilon$ has to be chosen differently for each species in order to take into account the large electron-ion mass ratio. For the simulations of this work, the choice $\epsilon_s=1.2(L_s/N)^{1.98}$ was reasonable for a sufficiently low number of markers. However, when the total number of markers is higher, it is likely that the choice of the $\epsilon_s$ parameter needs to be fine-tuned differently. With bad choices, temperature relaxation may fail, as illustrated in Fig.~\ref{fig:energy-jump-temperature-wrong}. 

\revision{Finally, while the particular discretization choice made in \eqref{time-discretization} conserves exactly the total energy and momentum, it does not say anything specific of the entropy evolution. To quantify the behaviour, the entropy functional \eqref{S} can be evaluated numerically and its value recorded during simulations. The entropy evolution for the test case of Section~\ref{subsec:electron-ion-velocity-relaxation} is depicted in Fig. \ref{dS-v-relaxation}. It illustrates that, after the equilibrium has been approximately reached, the numerically evaluated entropy expression decreases slightly. A potential approach to prevent non-monotonic entropy evolution is to modify the simple numerical integrator studied in the present work and adopt discrete-gradient \cite{Harten_Lax_Leer:1983, Gonzalez1996} integrators. In the multispecies case, the discrete equations of motion for the markers \eqref{discrete-evol} would be replaced by
\begin{align}\label{eq:d-evol-dgradient}
    \frac{\boldsymbol{v}_p^{n+1}-\boldsymbol{v}_p^n}{\Delta t}=\frac{1}{m_s} \sum_{\bar{s}} \sum_{\bar{p}} w_{\bar{p}} \nu_{s\bar{s}} \mathbb{Q}\left(\overline{\boldsymbol{\Gamma}_n^{n+1}(K, p, \bar{p})}\right) \cdot \overline{\boldsymbol{\Gamma}_n^{n+1}\left(S_\epsilon, p, \bar{p}\right)}
\end{align}
where $K$ is the kinetic energy function \eqref{discrete-energy} and $\overline{\boldsymbol{\Gamma}_n^{n+1}(A,p,\bar{p})}$ is modification of \eqref{gamma_h} according to
\begin{align}
\overline{\boldsymbol{\Gamma}_n^{n+1}(A, p, \bar{p})}=\left.\frac{1}{m_s w_p} \frac{\overline{\partial A}}{\partial \boldsymbol{v}_p}\right|_n ^{n+1}-\left.\frac{1}{m_{\bar{s}} w_{\bar{p}}} \frac{\overline{\partial A}}{\partial \boldsymbol{v}_{\bar{p}}}\right|_n ^{n+1}.
\end{align}
The discrete gradient appearing in the above equation, for which there are many existing options, is any such approximation of the derivative that satisfies the property
\begin{align}\label{eq:d-gradient-prop}
A^{n+1}-A^n = \left.\sum_s \sum_p\left(\boldsymbol{v}_p^{n+1}-\boldsymbol{v}_p^n\right) \cdot \overline{\frac{\partial A}{\partial \boldsymbol{v}_p}}\right|_n ^{n+1}.
\end{align}
If these modifications are adopted to the discrete equations of motion, then the entropy evolution will satisfy
\begin{align}
\frac{S_\epsilon^{n+1}-S_\epsilon^n}{\Delta t}&=\frac{1}{2} \sum_{s,\bar{s}} \sum_{p, \bar{p}} \nu_{s\bar{s}} w_p w_{\bar{p}} \overline{\bar{\Gamma}_n^{n+1}\left(S_\epsilon, p, \bar{p}\right)} \cdot \mathbb{Q}\left(\overline{\boldsymbol{\Gamma}_n^{n+1}(K, p, \bar{p})}\right) \cdot \overline{\Gamma_n^{n+1}\left(S_\epsilon, p, \bar{p}\right)} 
\geq 0,
\end{align}
where the $\geq$-sign follows from the fact that the operator $\mathbb{Q}$ is positive semidefinite. This modification will also retain the discrete energy and momentum conservation. 
\begin{figure}[!htb]
    \centering
    \includegraphics[width=0.5\textwidth]{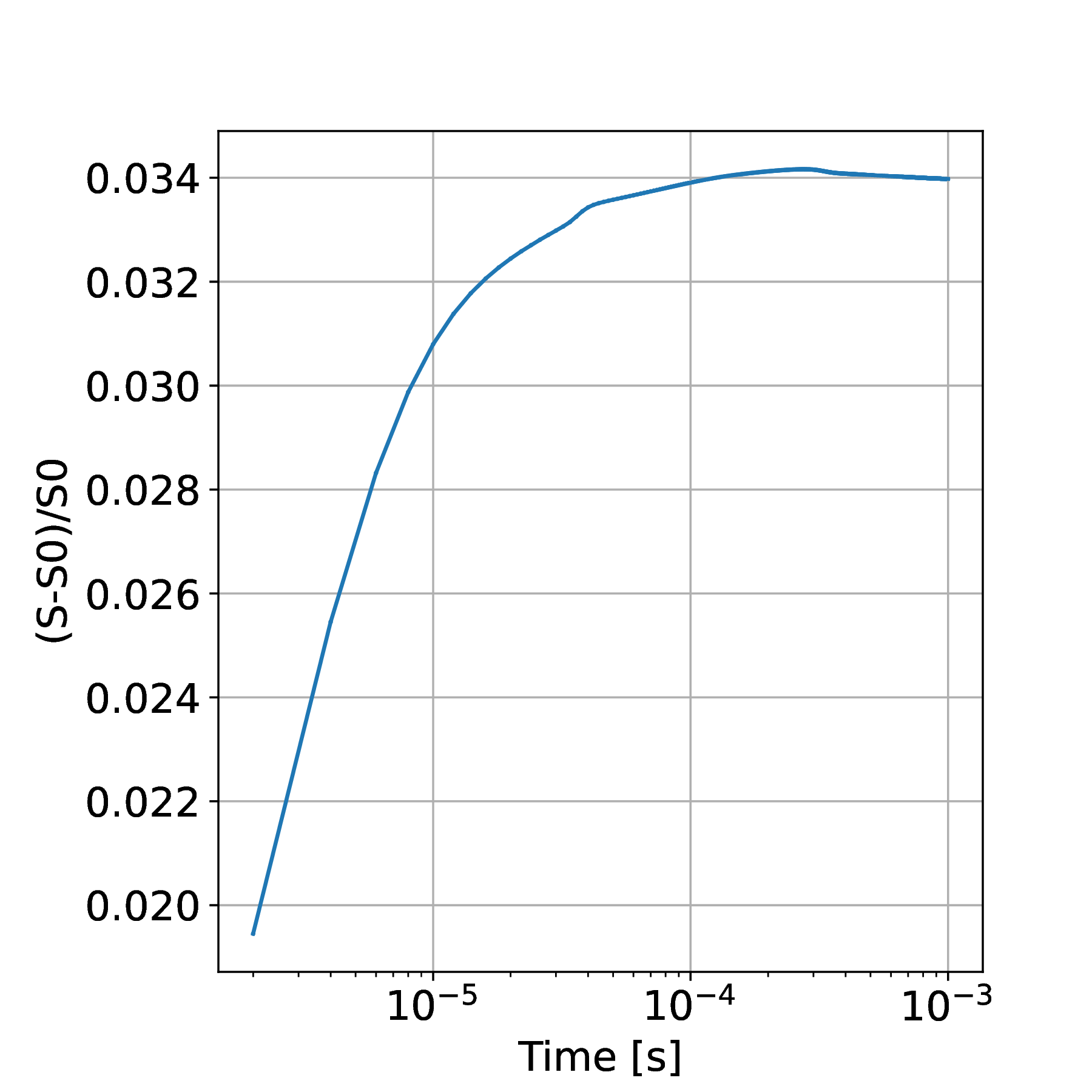}
    \caption{Entropy evolution for the velocity flow relaxation case presented in section \ref{subsec:electron-ion-velocity-relaxation}.} \label{dS-v-relaxation}
\end{figure}

The optimal choices for regularization parameters $\epsilon_s$ and the discrete gradient are left for future studies.
}

\section{Conclusions}\label{sec:conclusion}
This paper proposed a multispecies energy- and momentum-conserving numerical scheme for the nonlinear Landau collision operator based on arbitrarily weighted marker particles. The conservation properties were verified algebraically and the capability of the algorithm to reproduce velocity and temperature isotropization and thermalization was tested numerically against analytical theoretical predictions with good results. Open questions meriting further studies remain, in particular regarding the occasional inability of the solver to converge and the choice of the renormalization parameter. The use of discrete gradients to guarantee monotonic entropy production is also 
a needed subject for a future work. Nonetheless, the presented numerical scheme has clear potential to become a valid algorithm compatible with the structure-preserving particle-in-cell discretizations of the Vlasov--Maxwell system.

\section*{Acknowledgment}
The work of FZ was supported by the Academy of Finland grant no. 320058. The work of EH was supported by the Academy of Finland grant no. 315278.

\section*{Data availability statement}
The data that support the findings of this study are openly available in \url{https://github.com/zontafil/MultiSpeciesLandau} at \cite{github-multispecies}.

\bibliographystyle{unsrtnat}
\bibliography{bibfile}

\providecommand{\noopsort}[1]{#1}
\begin{thebibliography}{40}%
\makeatletter
\providecommand \@ifxundefined [1]{%
 \@ifx{#1\undefined}
}%
\providecommand \@ifnum [1]{%
 \ifnum #1\expandafter \@firstoftwo
 \else \expandafter \@secondoftwo
 \fi
}%
\providecommand \@ifx [1]{%
 \ifx #1\expandafter \@firstoftwo
 \else \expandafter \@secondoftwo
 \fi
}%
\providecommand \natexlab [1]{#1}%
\providecommand \enquote  [1]{``#1''}%
\providecommand \bibnamefont  [1]{#1}%
\providecommand \bibfnamefont [1]{#1}%
\providecommand \citenamefont [1]{#1}%
\providecommand \href@noop [0]{\@secondoftwo}%
\providecommand \href [0]{\begingroup \@sanitize@url \@href}%
\providecommand \@href[1]{\@@startlink{#1}\@@href}%
\providecommand \@@href[1]{\endgroup#1\@@endlink}%
\providecommand \@sanitize@url [0]{\catcode `\\12\catcode `\$12\catcode
  `\&12\catcode `\#12\catcode `\^12\catcode `\_12\catcode `\%12\relax}%
\providecommand \@@startlink[1]{}%
\providecommand \@@endlink[0]{}%
\providecommand \url  [0]{\begingroup\@sanitize@url \@url }%
\providecommand \@url [1]{\endgroup\@href {#1}{\urlprefix }}%
\providecommand \urlprefix  [0]{URL }%
\providecommand \Eprint [0]{\href }%
\providecommand \doibase [0]{https://doi.org/}%
\providecommand \selectlanguage [0]{\@gobble}%
\providecommand \bibinfo  [0]{\@secondoftwo}%
\providecommand \bibfield  [0]{\@secondoftwo}%
\providecommand \translation [1]{[#1]}%
\providecommand \BibitemOpen [0]{}%
\providecommand \bibitemStop [0]{}%
\providecommand \bibitemNoStop [0]{.\EOS\space}%
\providecommand \EOS [0]{\spacefactor3000\relax}%
\providecommand \BibitemShut  [1]{\csname bibitem#1\endcsname}%
\let\auto@bib@innerbib\@empty
\bibitem [{\citenamefont {{Squire}}\ \emph {et~al.}(2012)\citenamefont
  {{Squire}}, \citenamefont {{Qin}},\ and\ \citenamefont
  {{Tang}}}]{Squire-Qin-Tang-PIC:2012PhPl}%
  \BibitemOpen
  \bibfield  {author} {\bibinfo {author} {\bibfnamefont {J.}~\bibnamefont
  {{Squire}}}, \bibinfo {author} {\bibfnamefont {H.}~\bibnamefont {{Qin}}},\
  and\ \bibinfo {author} {\bibfnamefont {W.~M.}\ \bibnamefont {{Tang}}},\
  }\bibfield  {title} {\bibinfo {title} {{Geometric integration of the
  Vlasov-Maxwell system with a variational particle-in-cell scheme}},\ }\href
  {https://doi.org/10.1063/1.4742985} {\bibfield  {journal} {\bibinfo
  {journal} {Physics of Plasmas}\ }\textbf {\bibinfo {volume} {19}},\ \bibinfo
  {eid} {084501} (\bibinfo {year} {2012})},\ \Eprint
  {https://arxiv.org/abs/1401.6723} {arXiv:1401.6723} \BibitemShut {NoStop}%
\bibitem [{\citenamefont {{Evstatiev}}\ and\ \citenamefont
  {{Shadwick}}(2013)}]{Evstatiev-shadwick:2013JCoPh}%
  \BibitemOpen
  \bibfield  {author} {\bibinfo {author} {\bibfnamefont {E.~G.}\ \bibnamefont
  {{Evstatiev}}}\ and\ \bibinfo {author} {\bibfnamefont {B.~A.}\ \bibnamefont
  {{Shadwick}}},\ }\bibfield  {title} {\bibinfo {title} {{Variational
  formulation of particle algorithms for kinetic plasma simulations}},\ }\href
  {https://doi.org/10.1016/j.jcp.2013.03.006} {\bibfield  {journal} {\bibinfo
  {journal} {Journal of Computational Physics}\ }\textbf {\bibinfo {volume}
  {245}},\ \bibinfo {pages} {376} (\bibinfo {year} {2013})},\ \Eprint
  {https://arxiv.org/abs/1210.3743} {arXiv:1210.3743} \BibitemShut {NoStop}%
\bibitem [{\citenamefont {{Shadwick}}\ \emph {et~al.}(2014)\citenamefont
  {{Shadwick}}, \citenamefont {{Stamm}},\ and\ \citenamefont
  {{Evstatiev}}}]{Shadwick-Stamm-Evstatiev:2014PhPl}%
  \BibitemOpen
  \bibfield  {author} {\bibinfo {author} {\bibfnamefont {B.~A.}\ \bibnamefont
  {{Shadwick}}}, \bibinfo {author} {\bibfnamefont {A.~B.}\ \bibnamefont
  {{Stamm}}},\ and\ \bibinfo {author} {\bibfnamefont {E.~G.}\ \bibnamefont
  {{Evstatiev}}},\ }\bibfield  {title} {\bibinfo {title} {{Variational
  formulation of macro-particle plasma simulation algorithms}},\ }\href
  {https://doi.org/10.1063/1.4874338} {\bibfield  {journal} {\bibinfo
  {journal} {Physics of Plasmas}\ }\textbf {\bibinfo {volume} {21}},\ \bibinfo
  {eid} {055708} (\bibinfo {year} {2014})}\BibitemShut {NoStop}%
\bibitem [{\citenamefont {{Stamm}}\ \emph {et~al.}(2014)\citenamefont
  {{Stamm}}, \citenamefont {{Shadwick}},\ and\ \citenamefont
  {{Evstatiev}}}]{Stamm-Shadwick-Evstatiev:2014ITPS}%
  \BibitemOpen
  \bibfield  {author} {\bibinfo {author} {\bibfnamefont {A.~B.}\ \bibnamefont
  {{Stamm}}}, \bibinfo {author} {\bibfnamefont {B.}~\bibnamefont
  {{Shadwick}}},\ and\ \bibinfo {author} {\bibfnamefont {E.~G.}\ \bibnamefont
  {{Evstatiev}}},\ }\bibfield  {title} {\bibinfo {title} {{Variational
  Formulation of Macroparticle Models for Electromagnetic Plasma
  Simulations}},\ }\href {https://doi.org/10.1109/TPS.2014.2320461} {\bibfield
  {journal} {\bibinfo  {journal} {IEEE Transactions on Plasma Science}\
  }\textbf {\bibinfo {volume} {42}},\ \bibinfo {pages} {1747} (\bibinfo {year}
  {2014})}\BibitemShut {NoStop}%
\bibitem [{\citenamefont {{Xiao}}\ \emph {et~al.}(2015)\citenamefont {{Xiao}},
  \citenamefont {{Qin}}, \citenamefont {{Liu}}, \citenamefont {{He}},
  \citenamefont {{Zhang}},\ and\ \citenamefont
  {{Sun}}}]{Xiao-et-al-kinetic:2015PhPl}%
  \BibitemOpen
  \bibfield  {author} {\bibinfo {author} {\bibfnamefont {J.}~\bibnamefont
  {{Xiao}}}, \bibinfo {author} {\bibfnamefont {H.}~\bibnamefont {{Qin}}},
  \bibinfo {author} {\bibfnamefont {J.}~\bibnamefont {{Liu}}}, \bibinfo
  {author} {\bibfnamefont {Y.}~\bibnamefont {{He}}}, \bibinfo {author}
  {\bibfnamefont {R.}~\bibnamefont {{Zhang}}},\ and\ \bibinfo {author}
  {\bibfnamefont {Y.}~\bibnamefont {{Sun}}},\ }\bibfield  {title} {\bibinfo
  {title} {{Explicit high-order non-canonical symplectic particle-in-cell
  algorithms for Vlasov-Maxwell systems}},\ }\href
  {https://doi.org/10.1063/1.4935904} {\bibfield  {journal} {\bibinfo
  {journal} {Physics of Plasmas}\ }\textbf {\bibinfo {volume} {22}},\ \bibinfo
  {eid} {112504} (\bibinfo {year} {2015})},\ \Eprint
  {https://arxiv.org/abs/1510.06972} {arXiv:1510.06972} \BibitemShut {NoStop}%
\bibitem [{\citenamefont {{He}}\ \emph {et~al.}(2015)\citenamefont {{He}},
  \citenamefont {{Qin}}, \citenamefont {{Sun}}, \citenamefont {{Xiao}},
  \citenamefont {{Zhang}},\ and\ \citenamefont
  {{Liu}}}]{He-et-al-Hamiltonian-splitting:2015PhPl}%
  \BibitemOpen
  \bibfield  {author} {\bibinfo {author} {\bibfnamefont {Y.}~\bibnamefont
  {{He}}}, \bibinfo {author} {\bibfnamefont {H.}~\bibnamefont {{Qin}}},
  \bibinfo {author} {\bibfnamefont {Y.}~\bibnamefont {{Sun}}}, \bibinfo
  {author} {\bibfnamefont {J.}~\bibnamefont {{Xiao}}}, \bibinfo {author}
  {\bibfnamefont {R.}~\bibnamefont {{Zhang}}},\ and\ \bibinfo {author}
  {\bibfnamefont {J.}~\bibnamefont {{Liu}}},\ }\bibfield  {title} {\bibinfo
  {title} {{Hamiltonian time integrators for Vlasov-Maxwell equations}},\
  }\href {https://doi.org/10.1063/1.4938034} {\bibfield  {journal} {\bibinfo
  {journal} {Physics of Plasmas}\ }\textbf {\bibinfo {volume} {22}},\ \bibinfo
  {eid} {124503} (\bibinfo {year} {2015})},\ \Eprint
  {https://arxiv.org/abs/1505.06076} {arXiv:1505.06076} \BibitemShut {NoStop}%
\bibitem [{\citenamefont {{Qin}}\ \emph {et~al.}(2016)\citenamefont {{Qin}},
  \citenamefont {{Liu}}, \citenamefont {{Xiao}}, \citenamefont {{Zhang}},
  \citenamefont {{He}}, \citenamefont {{Wang}}, \citenamefont {{Sun}},
  \citenamefont {{Burby}}, \citenamefont {{Ellison}},\ and\ \citenamefont
  {{Zhou}}}]{Qin-et-al:2016NucFu}%
  \BibitemOpen
  \bibfield  {author} {\bibinfo {author} {\bibfnamefont {H.}~\bibnamefont
  {{Qin}}}, \bibinfo {author} {\bibfnamefont {J.}~\bibnamefont {{Liu}}},
  \bibinfo {author} {\bibfnamefont {J.}~\bibnamefont {{Xiao}}}, \bibinfo
  {author} {\bibfnamefont {R.}~\bibnamefont {{Zhang}}}, \bibinfo {author}
  {\bibfnamefont {Y.}~\bibnamefont {{He}}}, \bibinfo {author} {\bibfnamefont
  {Y.}~\bibnamefont {{Wang}}}, \bibinfo {author} {\bibfnamefont
  {Y.}~\bibnamefont {{Sun}}}, \bibinfo {author} {\bibfnamefont {J.~W.}\
  \bibnamefont {{Burby}}}, \bibinfo {author} {\bibfnamefont {L.}~\bibnamefont
  {{Ellison}}},\ and\ \bibinfo {author} {\bibfnamefont {Y.}~\bibnamefont
  {{Zhou}}},\ }\bibfield  {title} {\bibinfo {title} {{Canonical symplectic
  particle-in-cell method for long-term large-scale simulations of the
  Vlasov-Maxwell equations}},\ }\href
  {https://doi.org/10.1088/0029-5515/56/1/014001} {\bibfield  {journal}
  {\bibinfo  {journal} {Nuclear Fusion}\ }\textbf {\bibinfo {volume} {56}},\
  \bibinfo {eid} {014001} (\bibinfo {year} {2016})},\ \Eprint
  {https://arxiv.org/abs/1503.08334} {arXiv:1503.08334} \BibitemShut {NoStop}%
\bibitem [{\citenamefont {{Xiao}}\ \emph {et~al.}(2016)\citenamefont {{Xiao}},
  \citenamefont {{Qin}}, \citenamefont {{Morrison}}, \citenamefont {{Liu}},
  \citenamefont {{Yu}}, \citenamefont {{Zhang}},\ and\ \citenamefont
  {{He}}}]{Xiao-et-al-fluid:2016PhPl}%
  \BibitemOpen
  \bibfield  {author} {\bibinfo {author} {\bibfnamefont {J.}~\bibnamefont
  {{Xiao}}}, \bibinfo {author} {\bibfnamefont {H.}~\bibnamefont {{Qin}}},
  \bibinfo {author} {\bibfnamefont {P.~J.}\ \bibnamefont {{Morrison}}},
  \bibinfo {author} {\bibfnamefont {J.}~\bibnamefont {{Liu}}}, \bibinfo
  {author} {\bibfnamefont {Z.}~\bibnamefont {{Yu}}}, \bibinfo {author}
  {\bibfnamefont {R.}~\bibnamefont {{Zhang}}},\ and\ \bibinfo {author}
  {\bibfnamefont {Y.}~\bibnamefont {{He}}},\ }\bibfield  {title} {\bibinfo
  {title} {{Explicit high-order noncanonical symplectic algorithms for ideal
  two-fluid systems}},\ }\href {https://doi.org/10.1063/1.4967276} {\bibfield
  {journal} {\bibinfo  {journal} {Physics of Plasmas}\ }\textbf {\bibinfo
  {volume} {23}},\ \bibinfo {eid} {112107} (\bibinfo {year} {2016})},\ \Eprint
  {https://arxiv.org/abs/1606.07005} {arXiv:1606.07005} \BibitemShut {NoStop}%
\bibitem [{\citenamefont {{Kraus}}\ \emph {et~al.}(2017)\citenamefont
  {{Kraus}}, \citenamefont {{Kormann}}, \citenamefont {{Morrison}},\ and\
  \citenamefont {{Sonnendr{\"u}cker}}}]{Kraus-et-al:2017JPlPh}%
  \BibitemOpen
  \bibfield  {author} {\bibinfo {author} {\bibfnamefont {M.}~\bibnamefont
  {{Kraus}}}, \bibinfo {author} {\bibfnamefont {K.}~\bibnamefont {{Kormann}}},
  \bibinfo {author} {\bibfnamefont {P.~J.}\ \bibnamefont {{Morrison}}},\ and\
  \bibinfo {author} {\bibfnamefont {E.}~\bibnamefont {{Sonnendr{\"u}cker}}},\
  }\bibfield  {title} {\bibinfo {title} {{GEMPIC: geometric electromagnetic
  particle-in-cell methods}},\ }\href
  {https://doi.org/10.1017/S002237781700040X} {\bibfield  {journal} {\bibinfo
  {journal} {Journal of Plasma Physics}\ }\textbf {\bibinfo {volume} {83}},\
  \bibinfo {eid} {905830401} (\bibinfo {year} {2017})},\ \Eprint
  {https://arxiv.org/abs/1609.03053} {arXiv:1609.03053} \BibitemShut {NoStop}%
\bibitem [{\citenamefont {{Xiao}}\ \emph {et~al.}(2018)\citenamefont {{Xiao}},
  \citenamefont {{Qin}},\ and\ \citenamefont {{Liu}}}]{Xiao-et-al:2018PlST}%
  \BibitemOpen
  \bibfield  {author} {\bibinfo {author} {\bibfnamefont {J.}~\bibnamefont
  {{Xiao}}}, \bibinfo {author} {\bibfnamefont {H.}~\bibnamefont {{Qin}}},\ and\
  \bibinfo {author} {\bibfnamefont {J.}~\bibnamefont {{Liu}}},\ }\bibfield
  {title} {\bibinfo {title} {{Structure-preserving geometric particle-in-cell
  methods for Vlasov-Maxwell systems}},\ }\href
  {https://doi.org/10.1088/2058-6272/aac3d1} {\bibfield  {journal} {\bibinfo
  {journal} {Plasma Science and Technology}\ }\textbf {\bibinfo {volume}
  {20}},\ \bibinfo {eid} {110501} (\bibinfo {year} {2018})},\ \Eprint
  {https://arxiv.org/abs/1804.08823} {arXiv:1804.08823} \BibitemShut {NoStop}%
\bibitem [{\citenamefont {{Xiao}}\ and\ \citenamefont
  {{Qin}}(2021)}]{Xiao-Qin-6d-tokamak:2020arXiv}%
  \BibitemOpen
  \bibfield  {author} {\bibinfo {author} {\bibfnamefont {J.}~\bibnamefont
  {{Xiao}}}\ and\ \bibinfo {author} {\bibfnamefont {H.}~\bibnamefont {{Qin}}},\
  }\bibfield  {title} {\bibinfo {title} {{Explicit structure-preserving
  geometric particle-in-cell algorithm in curvilinear orthogonal coordinate
  systems and its applications to whole-device 6D kinetic simulations of
  tokamak physics}},\ }\href {https://doi.org/10.1088/2058-6272/abf125}
  {\bibfield  {journal} {\bibinfo  {journal} {Plasma Science and Technology}\
  }\textbf {\bibinfo {volume} {23}},\ \bibinfo {pages} {055102} (\bibinfo
  {year} {2021})},\ \Eprint {https://arxiv.org/abs/2004.08150}
  {arxiv:2004.08150} \BibitemShut {NoStop}%
\bibitem [{\citenamefont {Hirvijoki}\ \emph {et~al.}(2020)\citenamefont
  {Hirvijoki}, \citenamefont {Kormann},\ and\ \citenamefont
  {Zonta}}]{Hirvijoki_Kormann_Zonta:2020PoP}%
  \BibitemOpen
  \bibfield  {author} {\bibinfo {author} {\bibfnamefont {E.}~\bibnamefont
  {Hirvijoki}}, \bibinfo {author} {\bibfnamefont {K.}~\bibnamefont {Kormann}},\
  and\ \bibinfo {author} {\bibfnamefont {F.}~\bibnamefont {Zonta}},\ }\bibfield
   {title} {\bibinfo {title} {Subcycling of particle orbits in variational,
  geometric electromagnetic particle-in-cell methods},\ }\href
  {https://doi.org/10.1063/5.0006403} {\bibfield  {journal} {\bibinfo
  {journal} {Physics of Plasmas}\ }\textbf {\bibinfo {volume} {27}},\ \bibinfo
  {pages} {092506} (\bibinfo {year} {2020})},\ \Eprint
  {https://arxiv.org/abs/2002.10759} {arXiv:2002.10759} \BibitemShut {NoStop}%
\bibitem [{\citenamefont {Yoon}\ and\ \citenamefont {Chang}(2014)}]{Yoon2014}%
  \BibitemOpen
  \bibfield  {author} {\bibinfo {author} {\bibfnamefont {E.~S.}\ \bibnamefont
  {Yoon}}\ and\ \bibinfo {author} {\bibfnamefont {C.~S.}\ \bibnamefont
  {Chang}},\ }\bibfield  {title} {\bibinfo {title} {{A Fokker-Planck-Landau
  collision equation solver on two-dimensional velocity grid and its
  application to particle-in-cell simulation}},\ }\bibfield  {journal}
  {\bibinfo  {journal} {Physics of Plasmas}\ }\textbf {\bibinfo {volume}
  {21}},\ \href {https://doi.org/10.1063/1.4867359} {10.1063/1.4867359}
  (\bibinfo {year} {2014})\BibitemShut {NoStop}%
\bibitem [{\citenamefont {Hager}\ \emph {et~al.}(2016)\citenamefont {Hager},
  \citenamefont {Yoon}, \citenamefont {Ku}, \citenamefont {D'Azevedo},
  \citenamefont {Worley},\ and\ \citenamefont {Chang}}]{Hager2016}%
  \BibitemOpen
  \bibfield  {author} {\bibinfo {author} {\bibfnamefont {R.}~\bibnamefont
  {Hager}}, \bibinfo {author} {\bibfnamefont {E.~S.}\ \bibnamefont {Yoon}},
  \bibinfo {author} {\bibfnamefont {S.}~\bibnamefont {Ku}}, \bibinfo {author}
  {\bibfnamefont {E.~F.}\ \bibnamefont {D'Azevedo}}, \bibinfo {author}
  {\bibfnamefont {P.~H.}\ \bibnamefont {Worley}},\ and\ \bibinfo {author}
  {\bibfnamefont {C.~S.}\ \bibnamefont {Chang}},\ }\bibfield  {title} {\bibinfo
  {title} {{A fully non-linear multi-species Fokker-Planck-Landau collision
  operator for simulation of fusion plasma}},\ }\href
  {https://doi.org/10.1016/j.jcp.2016.03.064} {\bibfield  {journal} {\bibinfo
  {journal} {Journal of Computational Physics}\ }\textbf {\bibinfo {volume}
  {315}},\ \bibinfo {pages} {644} (\bibinfo {year} {2016})}\BibitemShut
  {NoStop}%
\bibitem [{\citenamefont {{Hirvijoki}}\ and\ \citenamefont
  {{Adams}}(2017)}]{Hirvijoki_Adams:2017PhPl}%
  \BibitemOpen
  \bibfield  {author} {\bibinfo {author} {\bibfnamefont {E.}~\bibnamefont
  {{Hirvijoki}}}\ and\ \bibinfo {author} {\bibfnamefont {M.~F.}\ \bibnamefont
  {{Adams}}},\ }\bibfield  {title} {\bibinfo {title} {{Conservative
  discretization of the Landau collision integral}},\ }\href
  {https://doi.org/10.1063/1.4979122} {\bibfield  {journal} {\bibinfo
  {journal} {Physics of Plasmas}\ }\textbf {\bibinfo {volume} {24}},\ \bibinfo
  {eid} {032121} (\bibinfo {year} {2017})},\ \Eprint
  {https://arxiv.org/abs/1611.07881} {arXiv:1611.07881} \BibitemShut {NoStop}%
\bibitem [{\citenamefont {Adams}\ \emph {et~al.}(2017)\citenamefont {Adams},
  \citenamefont {Hirvijoki}, \citenamefont {Knepley}, \citenamefont {Brown},
  \citenamefont {Isaac},\ and\ \citenamefont {Mills}}]{Adams_et_al:2017SIAM}%
  \BibitemOpen
  \bibfield  {author} {\bibinfo {author} {\bibfnamefont {M.~F.}\ \bibnamefont
  {Adams}}, \bibinfo {author} {\bibfnamefont {E.}~\bibnamefont {Hirvijoki}},
  \bibinfo {author} {\bibfnamefont {M.~G.}\ \bibnamefont {Knepley}}, \bibinfo
  {author} {\bibfnamefont {J.}~\bibnamefont {Brown}}, \bibinfo {author}
  {\bibfnamefont {T.}~\bibnamefont {Isaac}},\ and\ \bibinfo {author}
  {\bibfnamefont {R.}~\bibnamefont {Mills}},\ }\bibfield  {title} {\bibinfo
  {title} {Landau collision integral solver with adaptive mesh refinement on
  emerging architectures},\ }\href {https://doi.org/10.1137/17M1118828}
  {\bibfield  {journal} {\bibinfo  {journal} {SIAM Journal on Scientific
  Computing}\ }\textbf {\bibinfo {volume} {39}},\ \bibinfo {pages} {C452}
  (\bibinfo {year} {2017})},\ \Eprint {https://arxiv.org/abs/1702.08880}
  {arXiv:1702.08880} \BibitemShut {NoStop}%
\bibitem [{\citenamefont {{Shiroto}}\ and\ \citenamefont
  {{Sentoku}}(2019)}]{Shiroto-Sentoku:2019PhRvE}%
  \BibitemOpen
  \bibfield  {author} {\bibinfo {author} {\bibfnamefont {T.}~\bibnamefont
  {{Shiroto}}}\ and\ \bibinfo {author} {\bibfnamefont {Y.}~\bibnamefont
  {{Sentoku}}},\ }\bibfield  {title} {\bibinfo {title} {{Structure-preserving
  strategy for conservative simulation of the relativistic nonlinear
  Landau-Fokker-Planck equation}},\ }\href
  {https://doi.org/10.1103/PhysRevE.99.053309} {\bibfield  {journal} {\bibinfo
  {journal} {Physical Review E}\ }\textbf {\bibinfo {volume} {99}},\ \bibinfo
  {eid} {053309} (\bibinfo {year} {2019})},\ \Eprint
  {https://arxiv.org/abs/1902.07866} {arXiv:1902.07866} \BibitemShut {NoStop}%
\bibitem [{\citenamefont {Adams}\ \emph {et~al.}(2022)\citenamefont {Adams},
  \citenamefont {Brennan}, \citenamefont {Knepley},\ and\ \citenamefont
  {Wang}}]{adams2022a}%
  \BibitemOpen
  \bibfield  {author} {\bibinfo {author} {\bibfnamefont {M.~F.}\ \bibnamefont
  {Adams}}, \bibinfo {author} {\bibfnamefont {D.~P.}\ \bibnamefont {Brennan}},
  \bibinfo {author} {\bibfnamefont {M.~G.}\ \bibnamefont {Knepley}},\ and\
  \bibinfo {author} {\bibfnamefont {P.}~\bibnamefont {Wang}},\ }\bibfield
  {title} {\bibinfo {title} {{L}andau collision operator in the {CUDA}
  programming model applied to thermal quench plasmas},\ }in\ \href@noop {}
  {\emph {\bibinfo {booktitle} {IEEE International Parallel and Distributed
  Processing Symposium}}}\ (\bibinfo {year} {2022})\ \Eprint
  {https://arxiv.org/abs/2104.10000} {arXiv:2104.10000} \BibitemShut {NoStop}%
\bibitem [{\citenamefont {{Taitano}}\ \emph {et~al.}(2015)\citenamefont
  {{Taitano}}, \citenamefont {{Chac{\'o}n}}, \citenamefont {{Simakov}},\ and\
  \citenamefont {{Molvig}}}]{Taitano_et_al:2015JCP}%
  \BibitemOpen
  \bibfield  {author} {\bibinfo {author} {\bibfnamefont {W.~T.}\ \bibnamefont
  {{Taitano}}}, \bibinfo {author} {\bibfnamefont {L.}~\bibnamefont
  {{Chac{\'o}n}}}, \bibinfo {author} {\bibfnamefont {A.~N.}\ \bibnamefont
  {{Simakov}}},\ and\ \bibinfo {author} {\bibfnamefont {K.}~\bibnamefont
  {{Molvig}}},\ }\bibfield  {title} {\bibinfo {title} {{A mass, momentum, and
  energy conserving, fully implicit, scalable algorithm for the
  multi-dimensional, multi-species Rosenbluth-Fokker-Planck equation}},\ }\href
  {https://doi.org/10.1016/j.jcp.2015.05.025} {\bibfield  {journal} {\bibinfo
  {journal} {Journal of Computational Physics}\ }\textbf {\bibinfo {volume}
  {297}},\ \bibinfo {pages} {357} (\bibinfo {year} {2015})}\BibitemShut
  {NoStop}%
\bibitem [{\citenamefont {{Taitano}}\ \emph {et~al.}(2017)\citenamefont
  {{Taitano}}, \citenamefont {{Chac{\'o}n}},\ and\ \citenamefont
  {{Simakov}}}]{Taitano_et_al_2017JCoPh}%
  \BibitemOpen
  \bibfield  {author} {\bibinfo {author} {\bibfnamefont {W.~T.}\ \bibnamefont
  {{Taitano}}}, \bibinfo {author} {\bibfnamefont {L.}~\bibnamefont
  {{Chac{\'o}n}}},\ and\ \bibinfo {author} {\bibfnamefont {A.~N.}\ \bibnamefont
  {{Simakov}}},\ }\bibfield  {title} {\bibinfo {title} {{An
  equilibrium-preserving discretization for the nonlinear
  Rosenbluth-Fokker-Planck operator in arbitrary multi-dimensional geometry}},\
  }\href {https://doi.org/10.1016/j.jcp.2017.03.032} {\bibfield  {journal}
  {\bibinfo  {journal} {Journal of Computational Physics}\ }\textbf {\bibinfo
  {volume} {339}},\ \bibinfo {pages} {453} (\bibinfo {year}
  {2017})}\BibitemShut {NoStop}%
\bibitem [{\citenamefont {{Daniel}}\ \emph {et~al.}(2020)\citenamefont
  {{Daniel}}, \citenamefont {{Taitano}},\ and\ \citenamefont
  {{Chac{\'o}n}}}]{Daniel_et_al:2020CPC}%
  \BibitemOpen
  \bibfield  {author} {\bibinfo {author} {\bibfnamefont {D.}~\bibnamefont
  {{Daniel}}}, \bibinfo {author} {\bibfnamefont {W.~T.}\ \bibnamefont
  {{Taitano}}},\ and\ \bibinfo {author} {\bibfnamefont {L.}~\bibnamefont
  {{Chac{\'o}n}}},\ }\bibfield  {title} {\bibinfo {title} {{A fully implicit,
  scalable, conservative nonlinear relativistic Fokker-Planck 0D-2P solver for
  runaway electrons}},\ }\href {https://doi.org/10.1016/j.cpc.2020.107361}
  {\bibfield  {journal} {\bibinfo  {journal} {Computer Physics Communications}\
  }\textbf {\bibinfo {volume} {254}},\ \bibinfo {eid} {107361} (\bibinfo {year}
  {2020})},\ \Eprint {https://arxiv.org/abs/1902.10241} {arXiv:1902.10241}
  \BibitemShut {NoStop}%
\bibitem [{\citenamefont {{Takizuka}}\ and\ \citenamefont
  {{Abe}}(1977)}]{takizukaabe:1977}%
  \BibitemOpen
  \bibfield  {author} {\bibinfo {author} {\bibfnamefont {T.}~\bibnamefont
  {{Takizuka}}}\ and\ \bibinfo {author} {\bibfnamefont {H.}~\bibnamefont
  {{Abe}}},\ }\bibfield  {title} {\bibinfo {title} {{A Binary Collision Model
  for Plasma Simulation with a Particle Code}},\ }\href
  {https://doi.org/10.1016/0021-9991(77)90099-7} {\bibfield  {journal}
  {\bibinfo  {journal} {Journal of Computational Physics}\ }\textbf {\bibinfo
  {volume} {25}},\ \bibinfo {pages} {205} (\bibinfo {year} {1977})}\BibitemShut
  {NoStop}%
\bibitem [{\citenamefont {Miller}\ and\ \citenamefont
  {Combi}(1994)}]{Miller_Combi:1994}%
  \BibitemOpen
  \bibfield  {author} {\bibinfo {author} {\bibfnamefont {R.~H.}\ \bibnamefont
  {Miller}}\ and\ \bibinfo {author} {\bibfnamefont {M.~R.}\ \bibnamefont
  {Combi}},\ }\bibfield  {title} {\bibinfo {title} {{A Coulomb collision
  algorithm for weighted particle simulations}},\ }\href
  {https://doi.org/10.1029/94GL01835} {\bibfield  {journal} {\bibinfo
  {journal} {Geophysical Research Letters}\ }\textbf {\bibinfo {volume} {21}},\
  \bibinfo {pages} {1735} (\bibinfo {year} {1994})}\BibitemShut {NoStop}%
\bibitem [{\citenamefont {Wang}\ \emph {et~al.}(1996)\citenamefont {Wang},
  \citenamefont {Okamoto}, \citenamefont {Nakajima},\ and\ \citenamefont
  {Murakami}}]{WANG1996}%
  \BibitemOpen
  \bibfield  {author} {\bibinfo {author} {\bibfnamefont {W.}~\bibnamefont
  {Wang}}, \bibinfo {author} {\bibfnamefont {M.}~\bibnamefont {Okamoto}},
  \bibinfo {author} {\bibfnamefont {N.}~\bibnamefont {Nakajima}},\ and\
  \bibinfo {author} {\bibfnamefont {S.}~\bibnamefont {Murakami}},\ }\bibfield
  {title} {\bibinfo {title} {{Vector Implementation of Nonlinear Monte Carlo
  Coulomb Collisions}},\ }\href {https://doi.org/10.1006/jcph.1996.0204}
  {\bibfield  {journal} {\bibinfo  {journal} {Journal of Computational
  Physics}\ }\textbf {\bibinfo {volume} {128}},\ \bibinfo {pages} {209}
  (\bibinfo {year} {1996})}\BibitemShut {NoStop}%
\bibitem [{\citenamefont {{Nanbu}}(1997)}]{nanbu:1997}%
  \BibitemOpen
  \bibfield  {author} {\bibinfo {author} {\bibfnamefont {K.}~\bibnamefont
  {{Nanbu}}},\ }\bibfield  {title} {\bibinfo {title} {{Theory of cumulative
  small-angle collisions in plasmas}},\ }\href
  {https://doi.org/10.1103/PhysRevE.55.4642} {\bibfield  {journal} {\bibinfo
  {journal} {Physical Review E}\ }\textbf {\bibinfo {volume} {55}},\ \bibinfo
  {pages} {4642} (\bibinfo {year} {1997})}\BibitemShut {NoStop}%
\bibitem [{\citenamefont {Nanbu}\ and\ \citenamefont
  {Yonemura}(1998)}]{Nanbu_Yonemura:1998}%
  \BibitemOpen
  \bibfield  {author} {\bibinfo {author} {\bibfnamefont {K.}~\bibnamefont
  {Nanbu}}\ and\ \bibinfo {author} {\bibfnamefont {S.}~\bibnamefont
  {Yonemura}},\ }\bibfield  {title} {\bibinfo {title} {{Weighted Particles in
  Coulomb Collision Simulations Based on the Theory of a Cumulative Scattering
  Angle}},\ }\href {https://doi.org/10.1006/jcph.1998.6049} {\bibfield
  {journal} {\bibinfo  {journal} {Journal of Computational Physics}\ }\textbf
  {\bibinfo {volume} {145}},\ \bibinfo {pages} {639} (\bibinfo {year}
  {1998})}\BibitemShut {NoStop}%
\bibitem [{\citenamefont {{Carrillo}}\ \emph {et~al.}(2020)\citenamefont
  {{Carrillo}}, \citenamefont {{Hu}}, \citenamefont {{Wang}},\ and\
  \citenamefont {{Wu}}}]{Carrillo-et-al:2020}%
  \BibitemOpen
  \bibfield  {author} {\bibinfo {author} {\bibfnamefont {J.~A.}\ \bibnamefont
  {{Carrillo}}}, \bibinfo {author} {\bibfnamefont {J.}~\bibnamefont {{Hu}}},
  \bibinfo {author} {\bibfnamefont {L.}~\bibnamefont {{Wang}}},\ and\ \bibinfo
  {author} {\bibfnamefont {J.}~\bibnamefont {{Wu}}},\ }\bibfield  {title}
  {\bibinfo {title} {{A particle method for the homogeneous Landau equation}},\
  }\href {https://doi.org/10.1016/j.jcpx.2020.100066} {\bibfield  {journal}
  {\bibinfo  {journal} {Journal of Computational Physics: X}\ }\textbf
  {\bibinfo {volume} {7}},\ \bibinfo {eid} {100066} (\bibinfo {year} {2020})},\
  \Eprint {https://arxiv.org/abs/1910.03080} {arXiv:1910.03080} \BibitemShut
  {NoStop}%
\bibitem [{\citenamefont {Hirvijoki}(2021)}]{Hirvijoki2021}%
  \BibitemOpen
  \bibfield  {author} {\bibinfo {author} {\bibfnamefont {E.}~\bibnamefont
  {Hirvijoki}},\ }\bibfield  {title} {\bibinfo {title} {{Structure-preserving
  marker-particle discretizations of Coulomb collisions for particle-in-cell
  codes}},\ }\href {https://doi.org/10.1088/1361-6587/abe884} {\bibfield
  {journal} {\bibinfo  {journal} {Plasma Physics and Controlled Fusion}\
  }\textbf {\bibinfo {volume} {63}},\ \bibinfo {pages} {044003} (\bibinfo
  {year} {2021})},\ \Eprint {https://arxiv.org/abs/2012.07187}
  {arXiv:2012.07187} \BibitemShut {NoStop}%
\bibitem [{\citenamefont {Pusztay}(2022)}]{pusztay_dissertation}%
  \BibitemOpen
  \bibfield  {author} {\bibinfo {author} {\bibfnamefont {J.}~\bibnamefont
  {Pusztay}},\ }\emph {\bibinfo {title} {A Particle Basis Vlasov-Poisson-Landau
  Solver for Plasma Simulation in PETSc}},\ \href@noop {} {Ph.D. thesis},\
  \bibinfo  {school} {State University of New York at Buffalo} (\bibinfo {year}
  {2022})\BibitemShut {NoStop}%
\bibitem [{\citenamefont {Morrison}(1986)}]{Morrison:1986vw}%
  \BibitemOpen
  \bibfield  {author} {\bibinfo {author} {\bibfnamefont {P.~J.}\ \bibnamefont
  {Morrison}},\ }\bibfield  {title} {\bibinfo {title} {{A paradigm for joined
  Hamiltonian and dissipative systems}},\ }\href
  {https://doi.org/10.1016/0167-2789(86)90209-5} {\bibfield  {journal}
  {\bibinfo  {journal} {Physica D: Nonlinear Phenomena}\ }\textbf {\bibinfo
  {volume} {18}},\ \bibinfo {pages} {410} (\bibinfo {year} {1986})}\BibitemShut
  {NoStop}%
\bibitem [{\citenamefont {{Kraus}}\ and\ \citenamefont
  {{Hirvijoki}}(2017)}]{Kraus2017}%
  \BibitemOpen
  \bibfield  {author} {\bibinfo {author} {\bibfnamefont {M.}~\bibnamefont
  {{Kraus}}}\ and\ \bibinfo {author} {\bibfnamefont {E.}~\bibnamefont
  {{Hirvijoki}}},\ }\bibfield  {title} {\bibinfo {title} {{Metriplectic
  integrators for the Landau collision operator}},\ }\href
  {https://doi.org/10.1063/1.4998610} {\bibfield  {journal} {\bibinfo
  {journal} {Physics of Plasmas}\ }\textbf {\bibinfo {volume} {24}},\ \bibinfo
  {eid} {102311} (\bibinfo {year} {2017})},\ \Eprint
  {https://arxiv.org/abs/1707.01801} {arXiv:1707.01801} \BibitemShut {NoStop}%
\bibitem [{\citenamefont {Carrillo}\ \emph {et~al.}(2019)\citenamefont
  {Carrillo}, \citenamefont {Craig},\ and\ \citenamefont
  {Patacchini}}]{Carrillo2019}%
  \BibitemOpen
  \bibfield  {author} {\bibinfo {author} {\bibfnamefont {J.~A.}\ \bibnamefont
  {Carrillo}}, \bibinfo {author} {\bibfnamefont {K.}~\bibnamefont {Craig}},\
  and\ \bibinfo {author} {\bibfnamefont {F.~S.}\ \bibnamefont {Patacchini}},\
  }\bibfield  {title} {\bibinfo {title} {{A blob method for diffusion}},\
  }\href {https://doi.org/10.1007/s00526-019-1486-3} {\bibfield  {journal}
  {\bibinfo  {journal} {Calculus of Variations and Partial Differential
  Equations}\ }\textbf {\bibinfo {volume} {58}},\ \bibinfo {pages} {1}
  (\bibinfo {year} {2019})},\ \Eprint {https://arxiv.org/abs/1709.09195}
  {arXiv:1709.09195} \BibitemShut {NoStop}%
\bibitem [{\citenamefont {Press}\ \emph {et~al.}(1992)\citenamefont {Press},
  \citenamefont {Teukolsky}, \citenamefont {Vetterling},\ and\ \citenamefont
  {Flannery}}]{PresTeukVettFlan92}%
  \BibitemOpen
  \bibfield  {author} {\bibinfo {author} {\bibfnamefont {W.~H.}\ \bibnamefont
  {Press}}, \bibinfo {author} {\bibfnamefont {S.~A.}\ \bibnamefont
  {Teukolsky}}, \bibinfo {author} {\bibfnamefont {W.~T.}\ \bibnamefont
  {Vetterling}},\ and\ \bibinfo {author} {\bibfnamefont {B.~P.}\ \bibnamefont
  {Flannery}},\ }\href@noop {} {\emph {\bibinfo {title} {Numerical Recipes in
  C}}},\ \bibinfo {edition} {2nd}\ ed.\ (\bibinfo  {publisher} {Cambridge
  University Press},\ \bibinfo {address} {Cambridge, USA},\ \bibinfo {year}
  {1992})\BibitemShut {NoStop}%
\bibitem [{\citenamefont {Spitzer}\ and\ \citenamefont
  {Seeger}(1963)}]{Spitzer1963}%
  \BibitemOpen
  \bibfield  {author} {\bibinfo {author} {\bibfnamefont {L.}~\bibnamefont
  {Spitzer}}\ and\ \bibinfo {author} {\bibfnamefont {R.~J.}\ \bibnamefont
  {Seeger}},\ }\bibfield  {title} {\bibinfo {title} {{Physics of Fully Ionized
  Gases}},\ }\href {https://doi.org/10.1119/1.1969155} {\bibfield  {journal}
  {\bibinfo  {journal} {American Journal of Physics}\ }\textbf {\bibinfo
  {volume} {31}},\ \bibinfo {pages} {890} (\bibinfo {year} {1963})}\BibitemShut
  {NoStop}%
\bibitem [{\citenamefont {Hinton}\ and\ \citenamefont
  {Hazeltine}(1976)}]{Hinton1976}%
  \BibitemOpen
  \bibfield  {author} {\bibinfo {author} {\bibfnamefont {L.}~\bibnamefont
  {Hinton}}\ and\ \bibinfo {author} {\bibfnamefont {R.~D.}\ \bibnamefont
  {Hazeltine}},\ }\bibfield  {title} {\bibinfo {title} {{Theory of plasma
  transport in toroidal confinement systems}},\ }\href
  {https://journals.aps.org/rmp/abstract/10.1103/RevModPhys.48.239} {\bibfield
  {journal} {\bibinfo  {journal} {Reviews of Modern Physics}\ }\textbf
  {\bibinfo {volume} {48}},\ \bibinfo {pages} {239} (\bibinfo {year}
  {1976})}\BibitemShut {NoStop}%
\bibitem [{\citenamefont {Huba}(2013)}]{NRLFormulary}%
  \BibitemOpen
  \bibfield  {author} {\bibinfo {author} {\bibfnamefont {J.~D.}\ \bibnamefont
  {Huba}},\ }\href {http://wwwppd.nrl.navy.mil/nrlformulary/} {\emph {\bibinfo
  {title} {Plasma Physics}}}\ (\bibinfo  {publisher} {Naval Research
  Laboratory},\ \bibinfo {address} {Washington, DC},\ \bibinfo {year} {2013})\
  pp.\ \bibinfo {pages} {1--71}\BibitemShut {NoStop}%
\bibitem [{\citenamefont {Ichimaru}\ and\ \citenamefont
  {Rosenbluth}(1970)}]{Ichimaru1970}%
  \BibitemOpen
  \bibfield  {author} {\bibinfo {author} {\bibfnamefont {S.}~\bibnamefont
  {Ichimaru}}\ and\ \bibinfo {author} {\bibfnamefont {M.~N.}\ \bibnamefont
  {Rosenbluth}},\ }\bibfield  {title} {\bibinfo {title} {{Relaxation processes
  in plasmas with magnetic field. Temperature relaxations}},\ }\href
  {https://doi.org/10.1063/1.1692864} {\bibfield  {journal} {\bibinfo
  {journal} {Physics of Fluids}\ }\textbf {\bibinfo {volume} {13}},\ \bibinfo
  {pages} {2778} (\bibinfo {year} {1970})}\BibitemShut {NoStop}%
\bibitem [{\citenamefont {Harten}\ \emph {et~al.}(1983)\citenamefont {Harten},
  \citenamefont {Lax},\ and\ \citenamefont {Leer}}]{Harten_Lax_Leer:1983}%
  \BibitemOpen
  \bibfield  {author} {\bibinfo {author} {\bibfnamefont {A.}~\bibnamefont
  {Harten}}, \bibinfo {author} {\bibfnamefont {P.~D.}\ \bibnamefont {Lax}},\
  and\ \bibinfo {author} {\bibfnamefont {B.~v.}\ \bibnamefont {Leer}},\
  }\bibfield  {title} {\bibinfo {title} {On upstream differencing and
  godunov-type schemes for hyperbolic conservation laws},\ }\href
  {https://doi.org/10.1137/1025002} {\bibfield  {journal} {\bibinfo  {journal}
  {SIAM Review}\ }\textbf {\bibinfo {volume} {25}},\ \bibinfo {pages} {35}
  (\bibinfo {year} {1983})},\ \Eprint
  {https://arxiv.org/abs/https://doi.org/10.1137/1025002}
  {https://doi.org/10.1137/1025002} \BibitemShut {NoStop}%
\bibitem [{\citenamefont {Gonzalez}(1996)}]{Gonzalez1996}%
  \BibitemOpen
  \bibfield  {author} {\bibinfo {author} {\bibfnamefont {O.}~\bibnamefont
  {Gonzalez}},\ }\bibfield  {title} {\bibinfo {title} {{Time integration and
  discrete Hamiltonian systems}},\ }\href {https://doi.org/10.1007/BF02440162}
  {\bibfield  {journal} {\bibinfo  {journal} {Journal of Nonlinear Science}\
  }\textbf {\bibinfo {volume} {6}},\ \bibinfo {pages} {449} (\bibinfo {year}
  {1996})}\BibitemShut {NoStop}%
\bibitem [{\citenamefont {Zonta}\ \emph {et~al.}(2022)\citenamefont {Zonta},
  \citenamefont {Pusztay},\ and\ \citenamefont
  {Hirvijoki}}]{github-multispecies}%
  \BibitemOpen
  \bibfield  {author} {\bibinfo {author} {\bibfnamefont {F.}~\bibnamefont
  {Zonta}}, \bibinfo {author} {\bibfnamefont {J.~V.}\ \bibnamefont {Pusztay}},\
  and\ \bibinfo {author} {\bibfnamefont {E.}~\bibnamefont {Hirvijoki}},\ }\href
  {https://doi.org/10.5281/zenodo.7149196} {\bibinfo {title}
  {zontafil/multispecieslandau: Stable version}} (\bibinfo {year}
  {2022})\BibitemShut {NoStop}%
\end{thebibliography}%
\end{document}